\documentclass[aps,%
               prb,%
               reprint,%
               showpacs,%
               superscriptaddress,%
               longbibliography,%
               nofootinbib%
               ]{revtex4-2}

\usepackage[utf8]{inputenc}
\usepackage[usenames,dvipsnames]{xcolor}
\usepackage{graphicx}
\usepackage{amsfonts}
\usepackage{amssymb}
\usepackage{amsmath}
\usepackage{mathtools}
\usepackage{bm}
\usepackage{dsfont}
\usepackage{braket}

\graphicspath{{./figures/}}

\usepackage[pdfusetitle,%
            bookmarks=true,%
            colorlinks,%
            linkcolor=blue,%
            citecolor=blue,%
            urlcolor=blue%
            ]{hyperref}

\usepackage{tikz}
\usetikzlibrary{arrows, shapes, shapes.multipart, trees, positioning, backgrounds, fit, matrix, tikzmark}

\newcommand{\cre}[2]{{#1}^{\dagger}_{#2}}
\newcommand{\ann}[2]{{#1}^{\phantom{\dagger}}_{#2}}

\newcommand{\expval}[1]{\langle #1 \rangle}
\newcommand{\ham}{\hat{H}}
\newcommand{\bb}{\hat{b}^{\phantom\dagger}}
\newcommand{\bd}{\hat{b}^\dagger}
\newcommand{\db}{\hat{d}^{\phantom\dagger}}
\newcommand{\dd}{\hat{d}^\dagger}
\newcommand{\ddb}{\hat{d}^{(2)\phantom\dagger}}
\newcommand{\ddd}{\hat{d}^{(2)\dagger}}
\newcommand{\n}{\hat{n}}

\newcommand{\order}{\mathcal{O}}

\newcommand{\Fig}[1]{Figure\,{}\ref{#1}}
\newcommand{\fig}[1]{Fig.\,{}\ref{#1}}
\newcommand{\Subfig}[2]{Figure~\hyperref[fig:#1]{\ref*{fig:#1}(#2)}}
\newcommand{\subfig}[2]{Fig.\,\hyperref[fig:#1]{\ref*{fig:#1}(#2)}}
\newcommand{\Eq}[1]{Equation\,(\ref{#1})}
\newcommand{\eq}[1]{Eq.\,(\ref{#1})}

\newcommand{\eqs}[2]{Eqs.\,(\ref{#1},\ref{#2})}

\newcommand{\sect}[1]{Sec.\,\ref{#1}}

\newcommand{\TUM}{\affiliation{Technical University of Munich, TUM School of Natural Sciences, Physics Department, 85748 Garching, Germany}}
\newcommand{\MCQST}{\affiliation{Munich Center for Quantum Science and Technology (MCQST), Schellingstr. 4, 80799 M{\"u}nchen, Germany}}
\newcommand{\Harvard}{\affiliation{Department of Physics, Harvard University, Cambridge, MA 02138, USA}}

\begin{document}

%TC:ignore
\title{Dynamical Spectral Response of Fractonic Quantum Matter}
\author{Philip Zechmann}
\TUM \MCQST
\author{Julian Boesl}
\TUM \MCQST
\author{Johannes Feldmeier}
\Harvard
\author{Michael Knap}
\TUM \MCQST

\date{\today}

\begin{abstract}
Quantum many-body systems with fractonic excitations can realize fascinating phases of matter. Here, we study the low-energy excitations of a constrained Bose-Hubbard model in one dimension, which conserves the center of mass or, equivalently, the dipole moment in addition to the particle number. This model is known to realize fractonic phases, including a dipole Mott insulator, a dipole Luttinger liquid, and a metastable dipole supersolid. We use tensor network methods to compute spectral functions from the dynamical response of the system and verify predictions from low-energy field theories of the corresponding ground state phases. We demonstrate the existence of gapped excitations compatible with strong coupling results in a dipole Mott insulator, linear sound modes characteristic of a Luttinger liquid of dipoles, and soft quadratic modes at both zero and finite momenta in a supersolid state with charge density wave order and phase coherence at non-integer filling.
\end{abstract}

\maketitle
%TC:endignore
   
\section{Introduction} \label{sec:introduction}

Symmetries are ubiquitous in physics and constitute a major guiding principle in our understanding of quantum matter. Recently, systems with less conventional symmetries, particularly the conservation of higher moments of a global $U(1)$ charge, attracted great interest. These fractonic models~\cite{nandkishore:2019, pretko:2020, gromov:2022, chamon:2005, haah:2011, yoshida:2013, vijay:2015, vijay:2016, pretko:2017, pretko:2018b, pretko:2017a, williamson:2019} feature elementary excitations with constrained mobility, while collective correlated processes of multi-particle compounds can generate non-trivial dynamics. Inspired by an intriguing duality between fracton models and elasticity theory~\cite{pretko:2018, gromov:2019, kumar:2019, zhai:2019, radzihovsky:2020a, zhai:2021}, these models are candidates to realize quantum matter phases with rather unconventional properties, such as dipole superfluids and fracton condensates~\cite{pretko:2018, pretko:2018c, pretko:2019, yuan:2020, chen:2021, radzihovsky:2022, stahl:2023}. Recently, microscopic lattice models with dipole conservation were studied numerically, and the existence of fascinating ground-state phases in accordance with their respective low-energy effective theory could be identified~\cite{lake:2022, zechmann:2023, lake:2023, lake:2023a}. This paves a promising route towards the experimental realization of such phases of matter, especially as systems with dipole conservation have been successfully implemented in cold atom quantum simulators in the form of tilted Hubbard models~\cite{guardado-sanchez:2020, scherg:2021, zahn:2022, kohlert:2023}.

Dynamical probes in the form of spectral functions are a natural way to study the collective excitations of exotic phases. They reflect the dynamics of excitations above the ground state, characterizing the content of the excitation spectrum, and can be predicted from the respective low-energy effective theories. Connecting these predictions to microscopic models is, in general, a formidable task that requires substantial numerical effort. However, in one dimension, tensor network methods provide efficient numerical tools to study microscopic models. In this work, we compute spectral functions of the one-dimensional dipole-conserving Bose-Hubbard model, which hosts a rich ground-state phase diagram. We compare our numerical results to the corresponding low-energy effective field theories and provide a comprehensive numerical exploration of the excitation spectrum in a microscopic fracton model.

The work is structured as follows. In \sect{sec:model_phases}, we describe the microscopic model and discuss the phases it realizes, as well as their low-energy effective field theories. Spectral functions are introduced in \sect{sec:spectral_functions}, where we also present the corresponding sum rules. Numerical methods used to study the dynamical response are discussed in \sect{sec:methods}. Subsequently, we study the spectral function for dipole excitations in \sect{sec:dipole_spectra} and the dynamical structure factor in \sect{sec:density_spectra}. We conclude in \sect{sec:conclusion} with a summary of our findings and their implications on future experimental realizations. The sum rules of the spectral function are evaluated explicitly in the appendix.

\section{Model and groundstate phases} \label{sec:model_phases}

\begin{figure}
    \includegraphics[width=\columnwidth]{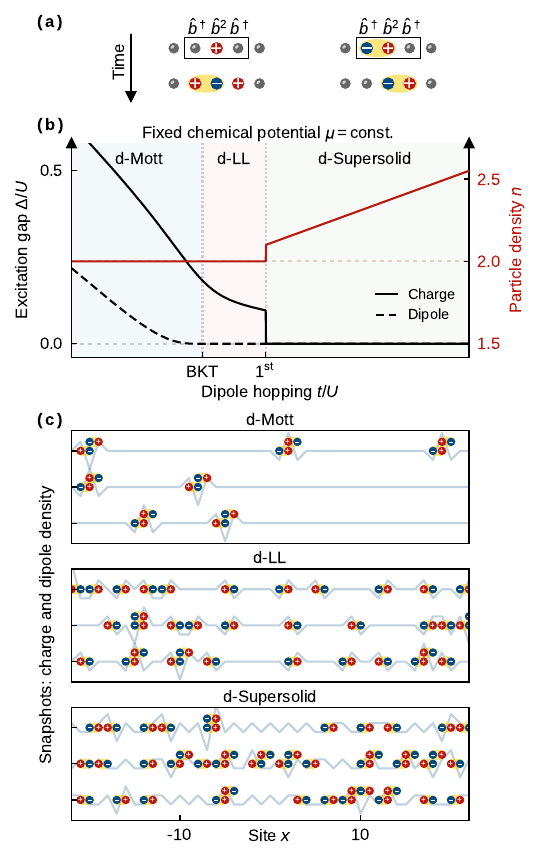}
    \caption{\label{fig:first}
        \textbf{Ground state phases.}
        (a)~Illustration of the dynamics generated by the dipole-conserving Bose-Hubbard model. Left: bosons can only hop in correlated moves, in which dipoles are emitted or absorbed when bosons hop. Right: the kinetic term can be seen as hopping of dipoles, i.e., boson-hole bound states.
        (b)~Ground state phases at fixed chemical potential $\mu/U = \text{const.}$ in a grand-canonical ensemble. By tuning the dipole hopping $t/U$, three phases characterized by the charge and the dipole gap can be realized: a dipole Mott insulator (d-Mott), a dipole Luttinger liquid (d-LL), and a dipole supersolid (d-Supersolid). Closing of the dipole gap accompanies the BKT transition from the d-Mott to the d-LL. The dipole gap vanishes at the first-order transition into the d-SS, where the boson density exhibits a jump.
        (c)~Selected snapshots of the charge (solid lines) and dipole density (symboles) in the respective phases extracted by sampling Fock configurations from the ground state MPS. 
    }
\end{figure}

Building upon previous studies~\cite{lake:2022, zechmann:2023, lake:2023}, we focus on a constrained Bose-Hubbard model in one dimension
\begin{equation} \label{eq:hamiltonian}
    \begin{split}
        \hat{H} = -t \sum_{j} (\cre{\hat{b}}{j}\ann{\hat{b}}{j+1}\ann{\hat{b}}{j+1}\cre{\hat{b}}{j+2} + \text{h.c.}) \\
        +\frac{U}{2}\sum_{j} \hat{n}_{j}(\hat{n}_{j}-1) - \mu\sum_{j} \hat{n}_{j}\,,
    \end{split}
\end{equation}
with a correlated hopping term of strength $t$, on-site interaction $U$, and chemical potential $\mu$. The kinetic term can be interpreted as nearest-neighbor hopping of dipoles $(\dd_{j}\db_{j+1} + \text{h.c.})$, where we defined the dipole creation and annihilation operators $\dd_{j} = \bd_{j}\bb_{j+1}$, and $\db_{j} = \bd_{j+1}\bb_{j}$, respectively. This model conserves both particle number $\hat{N} = \sum_j \hat{n}_j$ and the associated dipole moment, or center of mass, $\hat{P} = \sum_{j} j~\hat{n}_j$. As illustrated in \subfig{first}{a}, single bosons cannot hop independently but must emit or absorb dipoles under the dipole-conserving kinetic term, while dipoles show unrestricted mobility~\cite{nandkishore:2019, pretko:2020}.

Instead of the local particle density $n_j$, we may choose to describe the system via a local density of dipoles $n_{d,j}$, defined through~\cite{moudgalya:2021, feng:2022, zechmann:2023}
\begin{equation}\label{eq:density_mapping}
    \hat{n}_{d,j} = \sum_{l=0}^{j} (\n_{l}-n)\,,
\end{equation}
where $n=\braket{\hat N}/L$ is the average particle density. When cumulative charge fluctuations around their average satisfy an area law, the local dipole density is guaranteed to be bounded in the thermodynamic limit, rendering this mapping a powerful tool to understand the low-energy properties of the system~\cite{zechmann:2023}. For ground states of local Hamiltonians, a sufficient (but not necessary) condition for such area law fluctuations is the presence of a finite charge gap. In this case, the low-energy features of the system can be understood in terms of effective dipole degrees of freedom. We emphasize that the dipole operators $\dd_j$ introduced above are not, in general, canonical creation operators of the dipole density in \eq{eq:density_mapping}, i.e., $\hat{n}_{d,j} \neq \dd_{j}\db_{j}$. Instead, they entail effective interactions due to their non-trivial commutation relations $[\db_{j}, \dd_{j'}] \neq \delta_{j,j'}$.

The microscopic model \eq{eq:hamiltonian} realizes different exotic low-energy phases as depicted in \subfig{first}{b}~\cite{zechmann:2023, lake:2023}. At integer filling and small dipole hopping, the ground state is a gapped dipole Mott insulator with both finite charge- and dipole gap. Increasing the dipole hopping eventually leads to a Berezinskii-Kosterlitz-Thouless (BKT) transition into a dipole Luttinger liquid, accompanied by a closing dipole gap, while the charge gap remains finite. At large dipole hopping, the dipole Luttinger liquid exhibits an instability towards boson bunching due to the quartic scaling in bosonic operators of both the hopping and the interaction term. Away from integer filling, another intermediate phase arises. This can be illustrated in a grand canonical setting, where we fix the chemical potential and tune the dipole hopping; \subfig{first}{b}. There, a first-order transition into a compressible phase with a vanishing charge gap occurs, associated with a jump in the particle density. This phase is accompanied by coexisting off-diagonal long-range dipole order (ODLRO) and, for rational filling, charge density wave (CDW) order. Thus, it represents a `dipole supersolid'. In the thermodynamic limit, this phase is expected to acquire a finite charge gap due to lattice perturbations. However, on numerical accessible scales, this instability cannot be observed, demonstrating a remarkable robustness of this supersolid~\cite{zechmann:2023,lake:2023}. For irrational fillings, the system retains ODLRO but does not exhibit CDW order. Thus, a devil's staircase with incommensurate Lifschitz and commensurate metastable supersolid phases is realized in the dipolar Bose-Hubbard model when tuning the chemical potential.

A theoretical description of these low-energy phases is obtained with bosonization techniques~\cite{lake:2022, zechmann:2023, giamarchi:2004}, where a counting field $\phi(x)$ is introduced to express the charge density $n(x)$ as
\begin{equation} \label{eq:bosonization}
n(x) = \bigl[n-\frac{1}{\pi}\nabla \phi(x) \bigr]\sum_{m\in \mathbb{Z}}e^{2im\bigl(\pi n x-\phi(x)\bigr)}.
\end{equation}
We further introduce the canonically conjugate phase field $\theta(x)$, satisfying $[\partial_x \phi(x), \theta(x')]=-i\pi\delta(x-x')$. Analogously, we may bosonize the dipole degrees of freedom via the fields $\phi_d(x)$ and $\theta_d(x)$. At low energies, dipoles and charges are intimately connected through the continuum version of \eq{eq:density_mapping}, $\partial_x n_d(x) = n(x)$, thus linking the counting fields $\phi(x) = \partial_x \phi_d(x)$ according to \eq{eq:bosonization}. Integration by parts of the commutation relation directly reveals the conjugate phase field $[\partial_x \phi(x), \theta(x')] \overset{\text{P.I.}}{=} [\partial_x \phi_d(x), -\theta_d(x')]$, and we obtain the relations
\begin{equation}\label{eq:field_relations}
    \begin{split}
        \phi(x) &= \partial_x \phi_d(x) \\
        \partial_x \theta(x) &= -\theta_d(x) \,.
    \end{split}
\end{equation}
We may then formulate the field theory either in charge or dipole variables. 

In the presence of a finite charge gap at integer filling, it is convenient to work with dipoles as the low energy degrees of freedom, described by a sine-Gordon model
\begin{equation}\label{eq:sin-Gordon}
    \begin{split}
        H_{\text{SG}} = \frac{1}{2\pi}\int dx\, \bigg\{ \frac{u_d}{K_d}(\partial_x \phi_d)^2 + u_d&K_d(\partial_x \theta_d)^2 + \\
        &+ g \cos\left(2\phi_d\right) \bigg\} \,,
    \end{split}
\end{equation}
with Luttinger parameter $K_d$ and velocity $u_d$. At small $t/U$, the system realizes a Mott insulating phase for the dipoles. For $K_d < 2$, the lattice contribution $\cos(2\phi_d)$ is relevant and pins the counting field $\phi_d$, which gaps out the dipole degree of freedom and leads to exponentially decaying correlations. The system undergoes a BKT transition at $K_d=2$ (corresponding to some filling-dependent critical $t^*/U$), where the cosine becomes irrelevant, and the dipole gap closes. The low-energy effective theory is a Luttinger liquid of dipoles 
\begin{equation}\label{eq:luttinger_liquid}
    H_{\text{LL}} = \frac{1}{2\pi} \int dx \left\{ \frac{u_d}{K_d} (\partial_x \phi_d(x))^2 + u_d K_d (\partial_x \theta_d(x))^2 \right\} \,,
\end{equation}
with algebraic correlations and dynamical exponent $z=1$.
These two phases can also directly be discerned from snapshots of the charge density. \Subfig{first}{c} depicts selected snapshots of the charge density and the corresponding dipole density in the form of single pictorial dipoles with positive or negative charges. Fluctuations directly reveal qualitative features of the respective phase. In the dipole Mott insulator, only few deviations from the homogeneous state are visible and occur predominantly in dipole anti-dipole pairs. By contrast, for the dipole Luttinger liquid, dipole excitations become gapless, which is directly reflected in fluctuations comprised of single dipole excitations.

The previously discussed low-energy phases are realized at integer filling and preserve the translation invariance of the Hamiltonian. At noninteger filling, numerical evidence suggests a phase consistent with the phenomenology of a supersolid of dipoles at rational fillings~\cite{zechmann:2023, lake:2023}. The low energy physics of this phase is captured by a quantum Lifshitz model for \textit{gapless} charge degrees of freedom, given by
\begin{equation}\label{eq:lifshitz_model}
    H_{\text{Lif}} = \frac{1}{2\pi}\int dx\, \left[\frac{v}{K}(\partial_x\phi)^2 + vK(\partial_x^2\theta)^2\right] \,.
\end{equation}
Notably, the lowest order kinetic constribution to \eq{eq:lifshitz_model} compatible with dipole conservation is $(\partial_x^2\theta)^2$, invariant under $\theta(x) \rightarrow \theta(x) + a + bx$. This term originates from the long wavelength limit of dipole hopping, $\dd_j\db_{j+1}\sim e^{-i\partial_x^2\theta(x)}$, and induces a dynamical exponent $z=2$. The supersolid nature of \eq{eq:lifshitz_model} arises from a coexistence of ODLRO in the $\hat{d}$-operators and CDW order, as discussed in more detail in \sect{sec:dipole_spectra}. In \subfig{first}{c}, snapshots of the charge density exhibit regions where the periodic order (here, at filling $n=5/2$) is visible and superimposed with strong dipole fluctuations. Strictly speaking, the supersolid order can only be realized at rational fillings, whereas the Lifshitz model \eq{eq:lifshitz_model} should only describe the true ground state physics at irrational fillings. At rational filling, it is expected to be eventually unstable towards a finite charge gap due to lattice effects in the thermodynamic limit~\cite{zechmann:2023,lake:2023}. However, we note that the Lifshitz model appears to robustly capture the numerical simulations of \eq{eq:hamiltonian}~\cite{zechmann:2023,lake:2023} at rational fillings as well.

\section{Spectral functions and sum rules}  \label{sec:spectral_functions}
We study the excitation spectrum of our microscopic model \eq{eq:hamiltonian} by numerical evaluation of dynamical spectral functions. We introduce the dipole spectral function $A(\omega, k) = -\frac{1}{\pi} \text{Im} G^\text{r}(\omega, k)$, obtained from the retarded Greens function $G^\text{r}(t, k) = -i\Theta(t) \expval{[\db_{k}(t), \dd_{k}(0)]}$ for the dipole operators, as
\begin{equation}\label{eq:spectral_function}
    A(\omega, k) = \mathrm{Re} \int dt~ e^{i\omega t}~\langle [\db_{k}(t), \dd_{k}] \rangle\,. % \limits_{-\infty}^{\infty}
\end{equation}
Under spatial inversion, dipole creation operators map to dipole annihilation operators and vice versa, implying particle-hole symmetry for dipole excitations. For this reason, it is sufficient to only calculate either the particle or the hole spectrum in \eq{eq:spectral_function}. Additionally, we compute the dynamical structure factor
\begin{equation} \label{eq:dyn_structure_factor}
    S(\omega, k) = \int dt~ e^{i\omega t}~\langle \n_{k}(t) \n_{-k} \rangle\,, % \limits_{-\infty}^{\infty}
\end{equation}
describing the dynamical excitation spectrum of the density operator. In practice, the dynamical correlations entering \eqs{eq:spectral_function}{eq:dyn_structure_factor} are obtained by Fourier transformation of space-time correlations to momentum space,
\begin{equation}
    \begin{split}
    \langle \db_{k}(t) \dd_{k} \rangle\ &= 
    \frac{1}{L}\sum_{j, j'} e^{-ik(j-j')} \langle \db_{j}(t) \dd_{j'} \rangle \,, \\
    \langle \n_{k}(t) \n_{-k} \rangle\ &= 
    \frac{1}{L}\sum_{j, j'} e^{-ik(j-j')} \langle \n_{j}(t) \n_{j'} \rangle \,,
    \end{split}
\end{equation}
where expectation values are computed with respect to the ground state wave function.

The excitation spectrum can be characterized by its frequency moments, which we may evaluate analytically. The first moment of the dynamical structure factor is known as the f-sum rule
\begin{equation} \label{eq:canonical_fsum_rule}
    \int \frac{d\omega}{2\pi}\,\omega S(\omega, k) = f(k) \,.
\end{equation}
For non-constrained dynamics in the continuum, the f-sum rule for the density-density response is given by $f(k) = n k^2/2m$, where $n$ is the density and $m$ the mass~\cite{schwabl:2004}. Here, we present the results for our dipole-conserving lattice systems for the dipole and density operators. This can be done by calculating the corresponding commutators with the Hamiltonian. For the density operator, we obtain the f-sum rule
\begin{equation}\label{eq:fsum_density}
    \int \frac{d\omega}{2\pi} \omega\,S(\omega, k) = -8\sin^4(k/2)\,\langle \hat{T} \rangle \,,
\end{equation}
where $\expval{\hat{T}} = -2t\sum_j \expval{\dd_j\db_{j+1} + \text{h.c}}$ is the expectation value of the kinetic energy. Remarkably, we find the asymptotic momentum dependence $f(k) \sim k^4$, while for a regular, unconstrained hopping term, we would have $f(k) \sim k^2$. % (e.g., in the standard Bose-Hubbard model, one finds $\dots$).
Repeating this analysis for the dipole spectral function, we obtain
\begin{equation}\label{eq:fsum_dipole}
    \int \frac{d\omega}{2\pi} \omega\,A(\omega, k) = C_0 - C_1 \cos(k) + C_3 \cos^3(k) \,,
\end{equation}
where the coefficients $C_n$ contain various correlation functions evaluated with respect to the ground state. For explicit expressions, we refer to Appendix~\ref{apx:sum_rules}. For small momenta, the f-sum rule saturates as $f(k) \sim \text{const.} + k^2$, analogous to single-site boson operators in the conventional Bose-Hubbard model~\cite{freericks:2013}. \Eq{eq:fsum_dipole} is similar to this case, but the additional $\cos^3(k)$ term appears as a result of the non-canonical nature of the dipole operators.

\section{Numerical Approach}\label{sec:methods}
We use tensor network techniques to numerically compute the dynamical correlation function under controlled approximations. The well-established density matrix renormalization group (DMRG) algorithm~\cite{white:1992} provides an efficient way to variationally approximate the ground state wave function in matrix product state (MPS) representation. We truncate the local bosonic Hilbert space to a maximum occupation of $n_\text{max}=8$, which is high enough to ensure converged results. Our model exhibits particularly strong boundary effects for finite systems with open boundary conditions. Hence, we utilize the infinite system size version of DMRG (iDMRG)~\cite{vidal:2007} to eliminate the influence of the boundaries. Furthermore, our algorithms implement $U(1)$ particle density and dipole conservation. While constructing symmetric tensor networks, invariant under $U(1)$ transformations is well-understood~\cite{singh:2010,singh:2011}, we want to highlight how dipole conservation is realized. Generally, one introduces quantum numbers, or charges, e.g., $q_N$ for the particle density, on the legs of the tensors in a tensor network and demands that each tensor transforms such that the charges fulfill a Kirchhoff's law. In this sense, the dipole moment is just an additional leg charge, $q_P$, that we must keep track of. The key aspect here lies in the fact that translations do not commute with the dipole operator; hence, translating the MPS does not act trivially on the charges. In particular, we have to enforce the additional rule
\begin{equation}
    \left(q_N^{[n]}, q_P^{[n]}\right) \rightarrow \left(q_N^{[n]}, q_P^{[n]} + r q_N^{[n]}\right) \,,
\end{equation}
when translating the MPS by $r$ sites, where $n$ labels the tensor inside one unit cell. We note that dipole conservation severely restricts the connectivity of the variational space, and iDMGR might become non-ergodic and stuck in local minima~\cite{zaletel:2013}. To mitigate this issue, it is important to use a form of subspace expansion or density matrix mixing~\cite{white:2005, hubig:2015} to enlarge the reachable manifold for variational optimization.

After obtaining the groundstates with iDMRG, we compute the dynamical correlation function by applying the corresponding local operator, time-evolving the perturbed state, and eventually obtain the correlation function by computing the overlap. In the case of a translation invariant system, only a single time evolution is necessary for obtaining the spectral function. By contrast, when translation invariance is spontaneously broken, the time evolution has to be carried out on every sublattice.
%; see \apx{apx:spectral_fkt_trans_inv} for more details.
We use the time-dependent variational principle (TDVP) for MPS~\cite{haegeman:2011, haegeman:2016} to compute the time evolution. In the original 1-site formulation, TDVP cannot change the virtual bonds, and adjusting the charge sectors of the MPS tensors is not possible. Hence, we first use the 2-site variant of TDVP. Subsequently, when the state becomes sufficiently uniform, we switch to the 1-site variant due to the better scaling with the local Hilbert space dimension (which is significant in our case of bosons). As noted before, we want to avoid boundary effects using infinite MPS. For time evolution, we employ a hybrid scheme known as segment boundary conditions or window MPS, which describes an infinite chain effectively by a finite MPS~\cite{phien:2012, milsted:2013}. This is achieved by describing the dynamics inside a finite window embedded in an infinite MPS by treating the left and right parts as half-infinite groundstate MPS. Standard TDVP then realizes the time evolution but with non-trivial boundary environments, which describe the effective coupling to the half-infinite chains. These are obtained by computing the dominant left and right eigenvectors of the generalized transfer matrix.

The spectral function is subsequently computed by Fourier transformation in space and time. Due to time translation invariance, the negative time data can be obtained from the positive time data via complex conjugation. The energy resolution $\delta E \sim 1/\tau_{\max} $ is limited by the maximal evolution time accessible within the MPS approximation $\tau_\text{max}$, which in turn is restricted by the entanglement growth and controlled by the maximal bond dimension $\chi_\text{max}$. To avoid Gibbs ringing, we use artificial broadening by multiplying the correlator in the $[-\tau_\text{max}, \tau_\text{max}]$ interval with a Gaussian windowing function $\sim e^{-t^2/2\sigma^2}$ (with $\sigma \simeq \tau_\text{max}/2$), after applying linear prediction to extrapolate the data to longer times~\cite{press:1992,white:2008}.

The MPS formalism in the thermodynamic limit offers powerful tools to study elementary excitations. Isolated branches in the spectrum can be interpreted as quasiparticle excitations of the system and can be directly targeted by a variational ansatz~\cite{haegeman:2013}. In the spirit of generalizing the single-mode approximation, this ansatz describes quasiparticle excitations as a plane-wave superposition of local perturbations of the ground state
\begin{equation}\label{eq:qp_ansatz}
    |\Psi_p \rangle = \sum_n e^{inp} \Bigg[
    \begin{tikzpicture}
        [baseline={([yshift=0ex]current bounding box.center)},
         node distance=2mm,
         inner sep = .2mm,
         % on grid = true,
         tensor/.style={font=\footnotesize, rectangle, draw=black, fill=black!5, rounded corners=4pt, minimum size=16pt},
         leg/.style={font=\footnotesize, -}, %  thick,
        ]
        % Tensors
        \node [leg] (AL0) {$\dots$};
        \node [tensor] (AL1) [right=of AL0] {$A_L$};
        \node [tensor] (AL2) [right=of AL1] {$A_L$};
        \node [tensor] (B) [right=of AL2] {$B$};
        \node [tensor] (AR1) [right=of B] {$A_R$};
        \node [tensor] (AR2) [right=of AR1] {$A_R$};
        \node [leg] (AR3) [right=of AR2] {$\dots$};
        %
        % Connected legs
        \draw [leg] (AL0.east) -- (AL1.west);
        \draw [leg] (AL1.east) -- (AL2.west);
        \draw [leg] (AL2.east) -- (B.west);
        \draw [leg] (B.east) -- (AR1.west);
        \draw [leg] (AR1.east) -- (AR2.west);
        \draw [leg] (AR2.east) -- (AR3.west);
        % External Legs
        \node[below= 2mm of AL1, label=below:{\scriptsize $j_{n-1}$}] (s1) {};
        \node[below= 2mm of AL2, label=below:{\scriptsize $j_{n-1}$}] (s2) {};
        \node[below= 2mm of B, label=below:{\scriptsize $j_{n}$}] (s3) {};
        \node[below= 2mm of AR1, label=below:{\scriptsize $j_{n+1}$}] (s4) {};
        \node[below= 2mm of AR2, label=below:{\scriptsize $j_{+2}$}] (s5) {};
        \draw [leg] (s1.north) to (AL1.south);
        \draw [leg] (s2.north) to (AL2.south);
        \draw [leg] (s3.north) to (B.south);
        \draw [leg] (s4.north) to (AR1.south);
        \draw [leg] (s5.north) to (AR2.south);
    \end{tikzpicture}
    \Bigg] \,,
\end{equation}
where $A_L$ ($A_R$) are the translation invariant ground state MPS's left (right) isometric tensors, and $B$ represents the perturbation tensor to be optimized. Such states live in the tangent space of the ground state MPS, and appropriate gauge fixing allows for a parametrization for which overlaps can be computed efficiently~\cite{vanderstraeten:2019}. We have assumed a translation invariant MPS with a single-site unit cell, but the generalization to larger unit cells is straightforward. 

The quasiparticle ansatz describes the lowest possible state at a given momentum within the variational manifold for a given bond dimension. For gapped systems, this usually is the mode of interest. By contrast, for gapless systems, the ansatz may predict complex many-body excitations and the spectral weight has to be evaluated to assess the weight of these excitations in the targeted spectral function. For a given local operator $\hat{O}$, the spectral weight captured by the quasiparticle ansatz \eq{eq:qp_ansatz} is computed by evaluating overlaps with the ground state wavefunction
\begin{equation}
   Z_p = \Big|\langle \psi_0| \frac{1}{L} \sum_{n=0}^{L-1} e^{-ipn} \hat{O}_{n} |\Psi_p \rangle\Big|^2 \,.
\end{equation}

\begin{figure*}[t]
    \includegraphics[width=\textwidth]{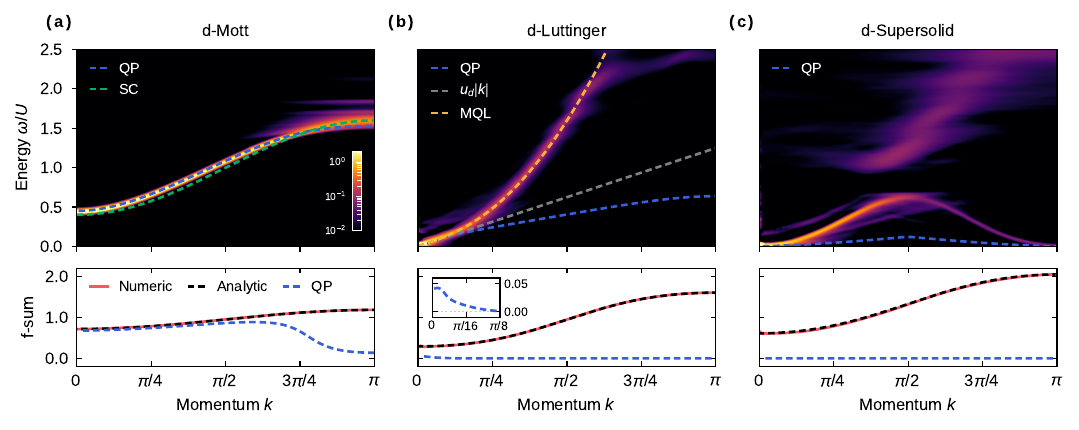}
    \caption{\label{fig:dipole_spectra}
        \textbf{Dipole spectral function.} 
        Characteristic excitations in
        (a)~the dipole Mott insulator at $n=2$ and $t/U=0.05$,
        (b)~the dipole Luttinger liquid at $n=2$ and $t/U=0.115$. and
        (c)~the dipole supersolid at $n=5/2$ and $t/U=0.05$. Top row: Dipole 
        spectral functions are compared to the MPS quasi-particle excitation ansatz (QP). For the dipole Mott insulator, additionally, the strong coupling (SC) result is shown, while for the Luttinger liquid, the Luttinger dispersion $\omega=u_d|k|$, and the dispersion of the massive quantum Lifshitz (MQL) model is illustrated.
        Bottom row: the f-sum rule, calculated from the numerically obtained spectrum (red) and the analytic prediction (black dashed). We also show the spectral weight of the lowest-lying quasi-particle excitation (QP), which is appreciable only in the gapped dipole Mott insulator phase, indicating that, in general, complex low-energy excitations can be formed in the dipole conserving Bose-Hubbard model, that not necessarily contribute to the pertinent spectral functions.
    }
\end{figure*}

\section{Dipole spectral function}\label{sec:dipole_spectra}

At low energies, dipoles constitute the relevant degrees of freedom, and the competition between their kinetic and the interaction energy determines the properties of the system. Therefore, we study the dynamical response of the system to perturbations that couple to the dipole operator $\dd_j$. Spatial inversion symmetry implies particle-hole symmetry for dipole excitations, as $\dd_{j} \rightarrow \db_{-j-1}$, and it is thus sufficient to study the particle excitation spectra. We numerically compute the spectral function for the three phases introduced above. In the dipole Mott and dipole Luttiner liquid phase, we fix the filling to be integer $n=2$ and choose values for $t/U$ below and above the critical $t^*/U = 0.113$~\cite{zechmann:2023}. \Subfig{dipole_spectra}{a,b} shows the resulting spectral functions. For the dipole supersolid phase, we fix a rational filling $n=5/2$ and choose an intermediate dipole hopping $t/U$; see \subfig{dipole_spectra}{c}.

\subsection{Dipole Mott insulator}
The low-energy physics of the dipole Mott insulator is described by the sine-Gordon model in \eq{eq:sin-Gordon}. Being the most relevant term, the cosine can be expanded around its minimum $\cos(2\phi_d) \sim \phi^2 $, which leads to gapped quadratic excitations
\begin{equation}
    \omega(k) =  \sqrt{(u_d k)^2 + \Delta_d^2} = \Delta_d + (u_d k)^2/2\Delta_d + \order(k^4) \,,
\end{equation}
with $\Delta_d = 4\pi g K_d u_d$. This is in accordance with the microscopic model, where, in the strong coupling limit $t/U\rightarrow 0$, dipole excitations on top of the Mott insulator behave essentially as free particles, and we find to lowest order the asymptotic dispersion
\begin{equation} \label{eq:strongcoupling}
    \lim_{t/U\rightarrow 0} \omega(k) = U - n(n+1) \, 2t\cos(k) \,.
\end{equation}
This agrees with the result for the conventional Bose-Hubbard model in the Mott phase~\cite{vanoosten:2001}, up to different prefactors resulting from the additional Bose factors of the correlated hopping term.

The spectral function, \subfig{dipole_spectra}{a}, exhibits a finite gap at $k=0$. It has a well-defined excitation branch at low momenta, which becomes broader at large momenta and eventually enters a multi-particle continuum. The dispersion obtained from the strong-coupling limit qualitatively captures the numerical results. Due to the gapped nature of the system, the quasi-particle ansatz shows good agreement with the spectral function, particularly at low momenta, where it exhausts the f-sum rule. In contrast, at large momenta, the reduced contribution of the lowest excited state indicates the existence of an excitation continuum.

As the dipole hopping increases, deviations from the strong coupling limit become substantial; see Appendix~\ref{apx:additional_data}. The dispersion tends to be increasingly linear near the $k=0$ point. Moreover, the dispersion obtained from the MPS quasiparticle ansatz lies below the frequencies with maximum weight in the spectral function. Here, more complex states are variationally found by the quasiparticle ansatz, which do not possess any relevant spectral weight.

\subsection{Dipole Luttinger liquid}
The dipole Luttinger liquid \eq{eq:luttinger_liquid} features gapless linear low-energy excitations
\begin{equation}\label{eq:luttinger_dispersion}
    \omega(k) = u_d |k| \,.
\end{equation}
Numerically, we directly obtain the velocity $u_d$ from the groundstate wave function via $u_d = K_d/\kappa_d \pi$, in which the Luttinger parameter is extracted from the algebraically decaying correlations $\expval{\dd_r \db_0} \sim r^{-1/2K_d}$, and the dipole compressibility from the finite size scaling of the dipole gap $\Delta_d(L)=\kappa_d^{-1}/L$.

The spectral function exhibits gapless excitations at $k=0$, \subfig{dipole_spectra}{b}, where most spectral weight is concentrated. At small momenta, the linear mode \eq{eq:luttinger_dispersion} with the Luttinger velocity $u_d$ agrees well with the excitation branch seen in the dynamical data. We emphasize that the linear dispersion is found in a relatively small momentum region. At larger momenta, it bends upwards before it enters a multi-particle continuum. Remarkably, a sharp quadratic mode with significant spectral weight exists outside the linear region. The quasiparticle ansatz does, in fact, correctly capture the linear dispersion at small momenta; however, at larger momenta, it fails. Generally, this is not surprising, as the variational states captured by the ansatz \eq{eq:qp_ansatz} are not guaranteed to work well for gapless systems, and there may well be a continuum of complex excitations with low spectral weight. This can be read off from the quasiparticle ansatz, as even in the $k\rightarrow 0$ limit, the full spectral sum rule is by far not exhausted.

Let us return to the well-defined quadratic mode observed in the spectrum. A natural candidate to explain this behavior is the quantum Lifhsitz model, and one might ask whether the dynamics away from the small momentum limit is determined by emergent \textit{Lifshitz-like} physics on the relevant time and length scales. We, therefore, consider a quantum Lifshitz model but with an additional mass term
\begin{equation}\label{eq:massive_lifshitz}
    H = \int dx \left\{ \frac{v}{2\pi} \left[ \frac{1}{K} (\partial_x \phi)^2 + K(\partial_x^2\theta)^2 \right] + g\phi^2\right\} \,.
\end{equation}
The Hamiltonian is expressed in charge degrees of freedom. This mass term results from the lowest-order contribution of the lattice correction $\cos(2\phi)$ present at integer filling, which is always relevant and ensures a finite charge gap. This model features gapless excitations with the dispersion
\begin{equation}
    \omega(k) = v |k| \sqrt{k^2 + 2\pi g K/v} \,.
\end{equation}
For small momenta $k \ll \sqrt{2\pi gK/v}$, the dipersion becomes linear $\omega(k) \sim |k|$. In this limit, we can map the theory to dipole degrees of freedom via \eq{eq:field_relations} and effectively obtain a Luttinger liquid with $u_d = \sqrt{2\pi gKv}$ and $K_d=\sqrt{Kv/2\pi g}$. Thus, the Luttinger liquid of dipoles is still the true low-energy theory under a renormalization group treatment, with dispersion $\omega(k) = u_d |k|$. However, for sufficiently large momenta $k \gg \sqrt{2\pi gK/v}$, when the charge gap becomes negligible, we effectively obtain a quantum Lifshitz model with quadratic dispersion $\omega(k) = v k^2$, compatible with the features observed in the numerical spectrum. At scales on which the system does not feel the presence of the small charge gap $\Delta_c \sim \frac{g}{Kv}$, it can thus realize Lifshitz physics. However, the charge gap becomes relevant at late times and large length scales, and the system flows to the Luttinger liquid fixed point, prohibiting a quadratic dispersion at small momenta. As shown in \subfig{dipole_spectra}{b}, the massive quantum Lifhsitz theory captures the observed excitations well using a single fitting parameter $K/v$, which is a non-universal function of the dipole hopping $t/U$.

\subsection{Dipole supersolid}
At rational filling, a phase compatible with the phenomenology of a supersolid arises. To illustrate this finding, let us fix the filling to be $n_0=p/q$, with p and q coprime integers. Reiterating the arguments of Ref.~\cite{seidel:2005}, the resulting ground state of a translation invariant Hamiltonian with both charge- and dipole-conservation is necessarily at least q-fold degenerate:
Let $\hat{T}$ be the translation operator, shifting by one lattice site, and $\hat{U} = e^{i2\pi/L \sum_j j \hat{n}}$ be the unitary transformation generated by the dipole operator $\hat{P}=\sum_j j \hat{n}$. The key point is as follows: Although both $\hat{U}$ and $\hat{T}$ comute with $\hat{H}$, $[\hat{U}, \ham] = [\hat{T}, \ham] = 0$, they do  not commute among themselves $[\hat{U}, \hat{T}] \neq 0$. From the definition of $\hat{U}$, it follows, $\hat{T}^{-1} \hat{U} \hat{T} = e^{i2\pi p/q} \hat{U}$, and hence commuting $\hat{U}$ and $\hat{T}$ results in an additional phases factor $\hat{U} \hat{T} = e^{i2\pi p/q} \hat{U}\hat{T}$. As [$\hat{U}, \ham]=0$, we can choose the eigenstates of the Hamiltonian to be simultaneous eigenstates of $\hat{U}$ and label them by the eigenvalue with respect to $\hat{U}$ as $|e^{i\phi}\rangle$. Moreover, because $[\hat{T}, \ham] = 0$, the translated state $\hat{T}|e^{i\phi}\rangle$ is also an eigenstate with the same energy, but its eigenvalue associated with $\hat{U}$ is altered $\hat{U}\hat{T}|e^{i\phi}\rangle = e^{i2\pi p/q}e^{i\phi}\hat{T}|e^{i\phi}\rangle$. Accordingly, translated eigenstates $\hat{T}^m|e^{i\phi}\rangle$ are orthogonal for $m<q$, and only after $q$ translations we arrive at the original eigenvalue $\hat{U} \hat{T}^{q} |e^{i\phi}\rangle = e^{i\phi} \hat{T}^{q} |e^{i\phi}\rangle$.
As a result, our system has at least $q$ degenerate ground states, which are connected by translations.

The low-energy theory is given by the quantum Lifshitz model presented in \eq{eq:lifshitz_model}. This theory features a modulated charge density with wave number $2\pi p/q$ for each of the groundstates
\begin{equation}
    \langle n(x) \rangle = n_0 + n_0 \sum_{m>0} \cos(2\pi x \, m \, p / q)~e^{-m^2 K/\alpha}\,,
\end{equation}
where $1/\alpha$ is the momentum cutoff necessary for UV regularization. Note that contributions from higher harmonics are strongly suppressed. The charge-density-wave order is captured by a finite order parameter
\begin{equation}
    \mathcal{O}_q = \frac{1}{L}\sum_j e^{i2\pi j p/q} \langle \hat{n}_j \rangle \neq 0\,.
\end{equation}
In addition, these ground states feature ODLRO in the dipole correlations $\expval{\dd_r\db_0} \xrightarrow{r\rightarrow \infty} \text{const.}$, and a finite superfluid stiffness~\cite{zechmann:2023}. They possess quadratic low-energy excitations
\begin{equation}
    \omega(k) = v k^2 \,.
\end{equation}
Due to the spontaneous breaking of translation symmetry, we expect $q$ of such soft modes located at momenta $k=2\pi m p /q \mod 2\pi$ with $m\in [0, q)$ and $m$ integer. These modes could, in principle, be thought of as softening of roton modes; however, due to the absence of a continuous phase transition into the dipole supersolid, there is no reason to expect precursors of them in any of the other phases.

The dipole spectral function at half-integer filling $n=5/2$ and dipole hopping $t/U=0.05$ exhibits the signatures of the dipole supersolid, \subfig{dipole_spectra}{c}. The spectrum has a vanishing excitation gap at both $k=0$ and $k=\pi$, as expected for $n=p/q$ with $q=2$. As the quantum Lifshitz theory predicts, these excitations are dispersing quadratically $\omega(k) \sim k^2$. CDW order leads to the additional soft mode at $k=\pi$, albeit with a very small spectral weight. 
The quasi-particle ansatz consistently fails to capture the observed spectrum, and the associated spectral weight has a vanishing contribution to the f-sum rule. We note that in the limit $k\rightarrow0$, the quasiparticle dispersion becomes compatible with a quadratic mode.
\begin{figure*}[t]
    \includegraphics[width=\textwidth]{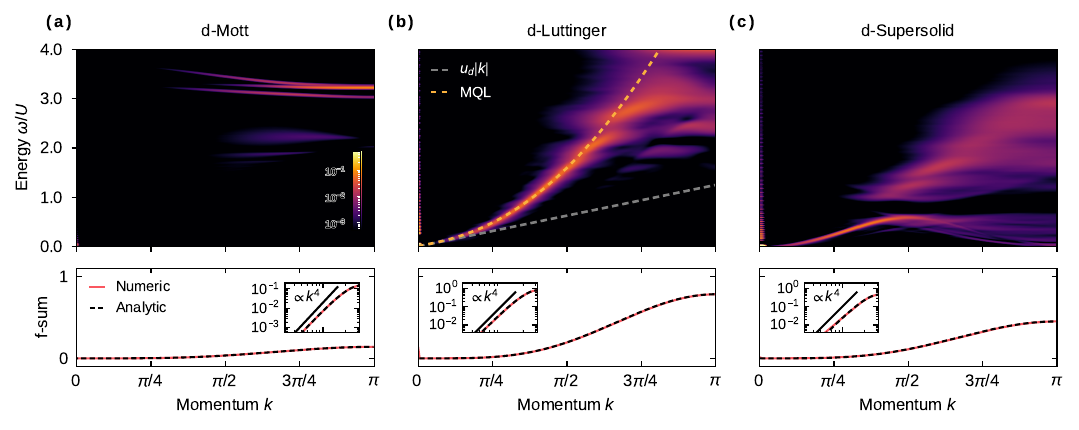}
    \caption{\label{fig:density_spectra}
        \textbf{Dynamical structure factor.}
        Top row: Dynamical structure factor in
        (a)~the dipole Mott,
        (b)~the dipole Luttinger liquid, and
        (c)~the dipole supersolid, for the same parameters as in \fig{fig:dipole_spectra}. We compare the spectrum of dipole Luttinger Liquid with the Luttinger dispersion $\omega = u_d |k|$ and the massive quantum Lifschitz (MQL) theory.
        Bottom row: f-sum rule of the dynamical structure factor. The insets highlight the $\sim k^4$ scaling of the f-sum rule at small momenta found in all three phases due to the conservation of the dipole moment.
    }
\end{figure*}

\section{Dynamical structure factor}\label{sec:density_spectra}
In addition to the dipole spectral function, we compute the dynamical structure factor defined in Eq.~\eqref{eq:dyn_structure_factor} for the different phases; see \fig{fig:density_spectra}. For all three phases, the spectral weight of the dynamical structure factor quickly vanishes for small momenta. This is simply a result of dipole conservation. As a consequence, charge transport is forbidden, leading to a vanishing DC conductivity. This is also reflected in the f-sum rule, which decays as $\sim k^4$ for small momenta, indicated in the insets where both the numerical data and analytic predictions are found to match precisely. 

A comparison of the structure factor and the dipole spectral function yields further insights. Most remarkably, for the dipole Luttinger liquid and the dipole supersolid, the low-energy excitation branches of both spectral functions coincide. This is in agreement with the analytical expectation from both the Luttinger liquid and Lifshitz low energy field theories. As those are quadratic theories, they predict unique gapless low energy modes that couple to both the dipole spectral function and the dynamical structure factor, albeit with different spectral weights.

By contrast, in the dipole Mott insulator, qualitative differences become visible. At strong coupling, the dipole spectral function creates mobile excitations on top of the dipole Mott insulator that follows the lattice dispersion, Eq.~\eqref{eq:strongcoupling}. By contrast, the dynamical structure factor probes collective density excitations that capture the fluctuations of the ground state. The dynamical structure factor, therefore, shows a weak gapped response at strong coupling.

\section{Conclusions and Outlook}\label{sec:conclusion}
In this work, we have numerically computed the excitation spectrum of the dipole-conserving Bose-Hubbard model. We compare the numerical results to predictions from low-energy effective field theories for the realized ground-state phases. We found sharply defined quasi-particle excitations in the dipole Mott insulator, in line with a strong coupling expansion. In the dipole Luttinger liquid phase, a linear gapless mode dominates at small momenta, as predicted by the low-energy theory. However, the mode bends to a quadratic dispersion for larger momenta. This can be understood from an effective quantum Lifshitz model with a charge gap and the coupling between charge and dipole degree of freedom. This intriguing finding suggests that, while the low-energy theory is an effective Luttinger liquid, Lifshitz physics is realized below a certain time and length scale, determined by the charge gap being small compared to the energy scale. This opens possibilities for finding unconventional signatures in dynamical probes on finite time scales accessible on experimental platforms. An interesting direction to further explore is how this observation influences far from equilibrium probes, such as the dynamics after a quench, i.e., by controllably adding particles or dipoles~\cite{boesl:2023}. Furthermore, we found a quadratic mode in the dipole supersolid phases, again compatible with the quantum Lifshitz theory. At fractional filling $n=p/q$, $q$ soft modes emerge at finite momenta due to the spontaneous breaking of translation symmetry, further endorsing the supersolid phenomenology.

Our finding may assist the experimental observation of fractonic quantum matter. Cold atom quantum simulators have demonstrated the realization of dipole-conserving models via linear potentials. Dynamical probes could offer a promising route to observe exotic phases, such as the dipole Luttinger liquid in one dimension, dipole condensation in higher dimensions, or even the additional soft modes in the symmetry broken phase at fractional filling. Our results provide a guiding principle for distinguishing these phases dynamically and corroborate the dynamics of dipoles and bosons in the ground states of fractonic models.

\begin{figure*}[t]
    \includegraphics[width=\textwidth]{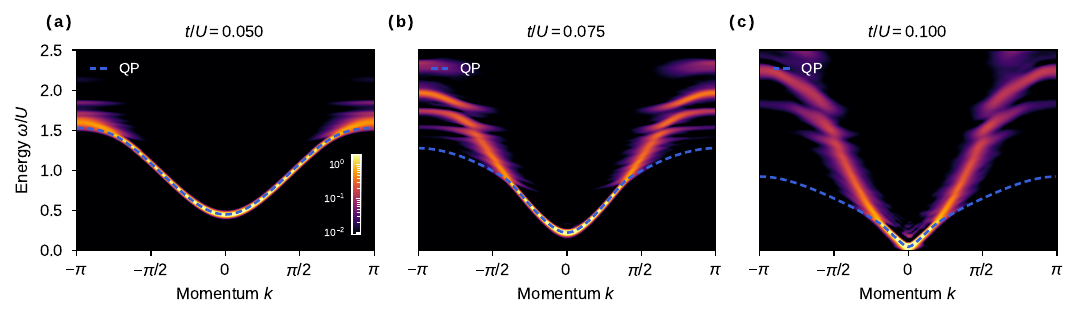}
    \caption{\label{fig:dipole_spectra_mott}
        \textbf{Dipole spectral function in the Mott insulating phase.}
        At integer filling $n=2$, the dipole hopping increases from left to right towards the BKT transition at $t/U=0.013$.
        % (a)~$t/U=0.025$,
        (a)~$t/U=0.050$,
        (b)~$t/U=0.075$, and
        (c)~$t/U=0.100$.
        Additionally, we show the lowest-lying excited states obtained from the MPS quasi-particle ansatz (QP).
    }
\end{figure*}
%

%TC:ignore
\begin{acknowledgments}
We thank Wilhelm Kadow, Anton Romen, and Laurens Vanderstraeten for insightful discussions.
We acknowledge support from the Deutsche Forschungsgemeinschaft (DFG, German Research Foundation) under Germany’s Excellence Strategy--EXC--2111--390814868 and DFG Grants No. KN1254/1-2, KN1254/2-1,  TRR 360 - 492547816, the European Research Council (ERC) under the European Union’s Horizon 2020 research and innovation programme (Grant Agreement No. 851161), as well as the Munich Quantum Valley, which is supported by the Bavarian state government with funds from the Hightech Agenda Bayern Plus.
JF acknowledges support by the Harvard Quantum Initiative.
Matrix product state simulations were performed using the TeNPy package~\cite{hauschild:2018}.
\end{acknowledgments}
%TC:endignore

\textit{Data and Code availability:} Numerical data and simulation codes are available on Zenodo upon reasonable request~\cite{zenodo}.

\appendix

\section{Spectral sum rules}\label{apx:sum_rules}
The moments of a spectral function of the form
\begin{equation}
    S(\omega) = \int dt~e^{i\omega t} \langle \hat{A}(t) \hat{B}(0) \rangle \,,
\end{equation}
can be computed exactly by calculating commutators of the corresponding operators with the Hamiltonian. For arbitrary integer power $n$, it holds that
\begin{equation}
    \int \frac{d\omega}{2\pi}~\omega^n\,S(\omega) = \langle [\hat{A}, \hat{H}]_n \hat{B} \rangle,
\end{equation}
with the nested commutator $[\hat{A}, \hat{B}]_n = [[\hat{A}, \hat{B}]_{n-1}, \hat{B}]$, and $[\hat{A}, \hat{B}]_{n=1} = [\hat{A}, \hat{B}]$. This follows directly from the Heisenberg equation of motion for $\hat{A}(t)$. In the following, we focus on the first moment $n=1$, known as the f-sum rule.

For the dynamical structure factor, we exploit inversion symmetry to recast this identity to the double commutator form
\begin{equation}
    \int \frac{d\omega}{2\pi}~\omega\,S(\omega, k) = \frac{1}{2} \langle [[\hat{n}_{k}, \hat{H}], \hat{n}_{-k}] \rangle \,,
\end{equation}
which can be evaluated exactly for the microscopic model. As pointed out before, the dipole operators do not fulfill canonical commutation relations, but we rather have
\begin{equation}
    \begin{split}
        [\db_j, \dd_{l}] &= \delta_{j,l} (\n_{j+1} - \n_{j})\,, \\
        [\db_j, \db_{l}] &= \delta_{j,l+1} \ddb_{j-1} - \delta_{j+1, l} \ddb_{j} \,, \\
        [\dd_j, \dd_{l}] &= \delta_{j+1,l} \ddd_{j} - \delta_{j, l+1} \ddd_{j-1} \,,
    \end{split}
\end{equation}
with the generalized dipole operator $\hat{d}_{j}^{\dagger (n)} = \bd_{j+n}\bb_{j}$. Additionally, it is useful to note the commutation relations with the density operator
\begin{equation}
    \begin{split}
    [\db_j, \n_l] = (\delta_{j,l} - \delta_{j+1,l}) \db_j \,, \\
    [\dd_j, \n_l] = (\delta_{j+1,l} - \delta_{j,l}) \dd_j \,.
    \end{split}
\end{equation}
The commutator with the density operator evaluates to
\begin{equation}
    [\n_j, \ham] = -t \left( \dd_{j}\db_{j+1} - 2\dd_{j-1}\db_{j} + \dd_{j-2}\db_{j-1} - \mathrm{h.c.} \right) \,.
\end{equation}
Note that the interactions commute with the number operator and do not contribute to the first moment. We obtain for the double commutator
\begin{equation}
     \langle [\hat{n}_k, \ham], \hat{n}_{-k}] \rangle = 32t \sin^4(k/2) \frac{1}{L} \sum_{j} \langle \dd_j \db_{j+1} \rangle \,,
\end{equation}
and accordingly the f-sum rule given in~\eq{eq:fsum_density}.

For the dipole spectral function, we have
\begin{equation}\label{eq:fsum_dipole_com}
    \int \frac{d\omega}{2\pi}~\omega\,A(\omega, k) = \langle [[\db_{k}, \hat{H}], \dd_{k}] \rangle \,.
\end{equation}
Due to the particle-hole symmetry of dipoles resulting from inversion symmetry, both particle and hole spectra contribute equally to the sum rule, and hence, the sum rule is fulfilled individually for them. In the main text, we only show particle spectra and, therefore, include a factor of $1/2$ on the right side of~\eq{eq:fsum_dipole_com}. The commutator of the dipole operator with the kinetic term gives
\begin{equation}
    \begin{split}
        [\db_j, \ham_t]
        % &= -t \sum_l [\db_j, \dd_{l}\db_{l+1}, \dd_{l+1}\db_{l}] \\
        & = -t \big\{ (\n_{j+1} - \n_j)(\db_{j+1} - \db_{j-1}) \\
                  + &(\dd_{j-2} + \dd_j)\ddb_{j-1} - (\dd_{j+2} + \dd_j)\ddb_{j} \big\} \,,
    \end{split}
\end{equation}
and with the interaction term (which, in contrast to the density operator,  does not vanish)
\begin{equation}
    [\db_j, \ham_U] = U \, (\n_j - \n_{j+1} + 1/2) \db_j
\end{equation}
Hence, in total, we obtain the sum rule
\begin{equation}
    \frac{1}{2} \langle [[\db_k, \ham], \dd_k] \rangle 
    = C_0 - C_1 \cos(k) + C_3 \cos^3(k) \,,
\end{equation}
where the coefficients $C_n$ contain various correlation functions evaluated on the ground state. 
For small momenta, the f-sum rule saturates as $f(k) \sim \text{const.} + k^2$, analogous to what is found in the standard Bose-Hubbard model for the bosonic creation and annihilation operators. The coefficients are given by
\begin{widetext}
    \begin{equation}
        \begin{aligned}
            & C_0 = \sum_j \left\{ 4t \langle \dd_j \db_{j+1} \rangle + U \left( \langle \dd_j \db_{j} \rangle - \langle \n_j \n_j \rangle + \langle \n_j \n_{j+1} \rangle\right) \right\} \,,\\
            &
            \begin{aligned}
                C_1 = \sum_j \left\{ U \langle \dd_j \db_{j+1} \rangle + t \left( 2\langle \dd_j\db_j \rangle + 2\langle \dd_j\db_{j+2} \rangle + 2\langle \n_j\n_{j+1} \rangle - \langle \n_j\n_{j+2} \rangle - \langle \n_j\n_j \rangle \right. \right. \\
                \left. + \left.3 \langle \ddd_j \ddb_{j+2} \rangle + 2 \langle \ddd_j \ddb_{j+1} \rangle - \langle \ddd_j \ddb_j \rangle \right) \right\} \,,\\
            \end{aligned} \\
            & C_3 = \sum_j 4t \langle \ddd_j \ddb_{j+2} \rangle \,.
        \end{aligned}
    \end{equation}
\end{widetext}
We numerically compute all the contributing correlation functions from the MPS ground state wavefunction when showing these results in the main text.

\newpage 
\section{Additional data}\label{apx:additional_data}

\Fig{fig:dipole_spectra_mott} presents supplementary data for the dipole spectral function in the Mott insulator for several dipole hopping strengths $t/U$.

\bibliography{references}

\begin{thebibliography}{57}%
\makeatletter
\providecommand \@ifxundefined [1]{%
 \@ifx{#1\undefined}
}%
\providecommand \@ifnum [1]{%
 \ifnum #1\expandafter \@firstoftwo
 \else \expandafter \@secondoftwo
 \fi
}%
\providecommand \@ifx [1]{%
 \ifx #1\expandafter \@firstoftwo
 \else \expandafter \@secondoftwo
 \fi
}%
\providecommand \natexlab [1]{#1}%
\providecommand \enquote  [1]{``#1''}%
\providecommand \bibnamefont  [1]{#1}%
\providecommand \bibfnamefont [1]{#1}%
\providecommand \citenamefont [1]{#1}%
\providecommand \href@noop [0]{\@secondoftwo}%
\providecommand \href [0]{\begingroup \@sanitize@url \@href}%
\providecommand \@href[1]{\@@startlink{#1}\@@href}%
\providecommand \@@href[1]{\endgroup#1\@@endlink}%
\providecommand \@sanitize@url [0]{\catcode `\\12\catcode `\$12\catcode
  `\&12\catcode `\#12\catcode `\^12\catcode `\_12\catcode `\%12\relax}%
\providecommand \@@startlink[1]{}%
\providecommand \@@endlink[0]{}%
\providecommand \url  [0]{\begingroup\@sanitize@url \@url }%
\providecommand \@url [1]{\endgroup\@href {#1}{\urlprefix }}%
\providecommand \urlprefix  [0]{URL }%
\providecommand \Eprint [0]{\href }%
\providecommand \doibase [0]{https://doi.org/}%
\providecommand \selectlanguage [0]{\@gobble}%
\providecommand \bibinfo  [0]{\@secondoftwo}%
\providecommand \bibfield  [0]{\@secondoftwo}%
\providecommand \translation [1]{[#1]}%
\providecommand \BibitemOpen [0]{}%
\providecommand \bibitemStop [0]{}%
\providecommand \bibitemNoStop [0]{.\EOS\space}%
\providecommand \EOS [0]{\spacefactor3000\relax}%
\providecommand \BibitemShut  [1]{\csname bibitem#1\endcsname}%
\let\auto@bib@innerbib\@empty
%</preamble>
\bibitem [{\citenamefont {Nandkishore}\ and\ \citenamefont
  {Hermele}(2019)}]{nandkishore:2019}%
  \BibitemOpen
  \bibfield  {author} {\bibinfo {author} {\bibfnamefont {R.~M.}\ \bibnamefont
  {Nandkishore}}\ and\ \bibinfo {author} {\bibfnamefont {M.}~\bibnamefont
  {Hermele}},\ }\bibfield  {title} {\bibinfo {title} {Fractons},\ }\href
  {https://doi.org/10.1146/annurev-conmatphys-031218-013604} {\bibfield
  {journal} {\bibinfo  {journal} {Annu. Rev. Condens. Matter Phys.}\ }\textbf
  {\bibinfo {volume} {10}},\ \bibinfo {pages} {295} (\bibinfo {year}
  {2019})}\BibitemShut {NoStop}%
\bibitem [{\citenamefont {Pretko}\ \emph {et~al.}(2020)\citenamefont {Pretko},
  \citenamefont {Chen},\ and\ \citenamefont {You}}]{pretko:2020}%
  \BibitemOpen
  \bibfield  {author} {\bibinfo {author} {\bibfnamefont {M.}~\bibnamefont
  {Pretko}}, \bibinfo {author} {\bibfnamefont {X.}~\bibnamefont {Chen}},\ and\
  \bibinfo {author} {\bibfnamefont {Y.}~\bibnamefont {You}},\ }\bibfield
  {title} {\bibinfo {title} {Fracton phases of matter},\ }\href
  {https://doi.org/10.1142/S0217751X20300033} {\bibfield  {journal} {\bibinfo
  {journal} {Int. J. Mod. Phys. A}\ }\textbf {\bibinfo {volume} {35}},\
  \bibinfo {pages} {2030003} (\bibinfo {year} {2020})}\BibitemShut {NoStop}%
\bibitem [{\citenamefont {Gromov}\ and\ \citenamefont
  {Radzihovsky}(2022)}]{gromov:2022}%
  \BibitemOpen
  \bibfield  {author} {\bibinfo {author} {\bibfnamefont {A.}~\bibnamefont
  {Gromov}}\ and\ \bibinfo {author} {\bibfnamefont {L.}~\bibnamefont
  {Radzihovsky}},\ }\href@noop {} {\bibinfo {title} {Fracton {{Matter}}}}
  (\bibinfo {year} {2022}),\ \Eprint {https://arxiv.org/abs/2211.05130}
  {arxiv:2211.05130} \BibitemShut {NoStop}%
\bibitem [{\citenamefont {Chamon}(2005)}]{chamon:2005}%
  \BibitemOpen
  \bibfield  {author} {\bibinfo {author} {\bibfnamefont {C.}~\bibnamefont
  {Chamon}},\ }\bibfield  {title} {\bibinfo {title} {Quantum {{Glassiness}} in
  {{Strongly Correlated Clean Systems}}: {{An Example}} of {{Topological
  Overprotection}}},\ }\href {https://doi.org/10.1103/PhysRevLett.94.040402}
  {\bibfield  {journal} {\bibinfo  {journal} {Phys. Rev. Lett.}\ }\textbf
  {\bibinfo {volume} {94}},\ \bibinfo {pages} {040402} (\bibinfo {year}
  {2005})}\BibitemShut {NoStop}%
\bibitem [{\citenamefont {Haah}(2011)}]{haah:2011}%
  \BibitemOpen
  \bibfield  {author} {\bibinfo {author} {\bibfnamefont {J.}~\bibnamefont
  {Haah}},\ }\bibfield  {title} {\bibinfo {title} {Local stabilizer codes in
  three dimensions without string logical operators},\ }\href
  {https://doi.org/10.1103/PhysRevA.83.042330} {\bibfield  {journal} {\bibinfo
  {journal} {Phys. Rev. A}\ }\textbf {\bibinfo {volume} {83}},\ \bibinfo
  {pages} {042330} (\bibinfo {year} {2011})}\BibitemShut {NoStop}%
\bibitem [{\citenamefont {Yoshida}(2013)}]{yoshida:2013}%
  \BibitemOpen
  \bibfield  {author} {\bibinfo {author} {\bibfnamefont {B.}~\bibnamefont
  {Yoshida}},\ }\bibfield  {title} {\bibinfo {title} {Exotic topological order
  in fractal spin liquids},\ }\href
  {https://doi.org/10.1103/PhysRevB.88.125122} {\bibfield  {journal} {\bibinfo
  {journal} {Phys. Rev. B}\ }\textbf {\bibinfo {volume} {88}},\ \bibinfo
  {pages} {125122} (\bibinfo {year} {2013})}\BibitemShut {NoStop}%
\bibitem [{\citenamefont {Vijay}\ \emph {et~al.}(2015)\citenamefont {Vijay},
  \citenamefont {Haah},\ and\ \citenamefont {Fu}}]{vijay:2015}%
  \BibitemOpen
  \bibfield  {author} {\bibinfo {author} {\bibfnamefont {S.}~\bibnamefont
  {Vijay}}, \bibinfo {author} {\bibfnamefont {J.}~\bibnamefont {Haah}},\ and\
  \bibinfo {author} {\bibfnamefont {L.}~\bibnamefont {Fu}},\ }\bibfield
  {title} {\bibinfo {title} {A new kind of topological quantum order: {{A}}
  dimensional hierarchy of quasiparticles built from stationary excitations},\
  }\href {https://doi.org/10.1103/PhysRevB.92.235136} {\bibfield  {journal}
  {\bibinfo  {journal} {Phys. Rev. B}\ }\textbf {\bibinfo {volume} {92}},\
  \bibinfo {pages} {235136} (\bibinfo {year} {2015})}\BibitemShut {NoStop}%
\bibitem [{\citenamefont {Vijay}\ \emph {et~al.}(2016)\citenamefont {Vijay},
  \citenamefont {Haah},\ and\ \citenamefont {Fu}}]{vijay:2016}%
  \BibitemOpen
  \bibfield  {author} {\bibinfo {author} {\bibfnamefont {S.}~\bibnamefont
  {Vijay}}, \bibinfo {author} {\bibfnamefont {J.}~\bibnamefont {Haah}},\ and\
  \bibinfo {author} {\bibfnamefont {L.}~\bibnamefont {Fu}},\ }\bibfield
  {title} {\bibinfo {title} {Fracton topological order, generalized lattice
  gauge theory, and duality},\ }\href
  {https://doi.org/10.1103/PhysRevB.94.235157} {\bibfield  {journal} {\bibinfo
  {journal} {Phys. Rev. B}\ }\textbf {\bibinfo {volume} {94}},\ \bibinfo
  {pages} {235157} (\bibinfo {year} {2016})}\BibitemShut {NoStop}%
\bibitem [{\citenamefont {Pretko}(2017{\natexlab{a}})}]{pretko:2017}%
  \BibitemOpen
  \bibfield  {author} {\bibinfo {author} {\bibfnamefont {M.}~\bibnamefont
  {Pretko}},\ }\bibfield  {title} {\bibinfo {title} {Subdimensional particle
  structure of higher rank {$U(1)$} spin liquids},\ }\href
  {https://doi.org/10.1103/PhysRevB.95.115139} {\bibfield  {journal} {\bibinfo
  {journal} {Phys. Rev. B}\ }\textbf {\bibinfo {volume} {95}},\ \bibinfo
  {pages} {115139} (\bibinfo {year} {2017}{\natexlab{a}})}\BibitemShut
  {NoStop}%
\bibitem [{\citenamefont {Pretko}\ and\ \citenamefont
  {Radzihovsky}(2018{\natexlab{a}})}]{pretko:2018b}%
  \BibitemOpen
  \bibfield  {author} {\bibinfo {author} {\bibfnamefont {M.}~\bibnamefont
  {Pretko}}\ and\ \bibinfo {author} {\bibfnamefont {L.}~\bibnamefont
  {Radzihovsky}},\ }\bibfield  {title} {\bibinfo {title} {Symmetry-{{Enriched
  Fracton Phases}} from {{Supersolid Duality}}},\ }\href
  {https://doi.org/10.1103/PhysRevLett.121.235301} {\bibfield  {journal}
  {\bibinfo  {journal} {Phys. Rev. Lett.}\ }\textbf {\bibinfo {volume} {121}},\
  \bibinfo {pages} {235301} (\bibinfo {year} {2018}{\natexlab{a}})}\BibitemShut
  {NoStop}%
\bibitem [{\citenamefont {Pretko}(2017{\natexlab{b}})}]{pretko:2017a}%
  \BibitemOpen
  \bibfield  {author} {\bibinfo {author} {\bibfnamefont {M.}~\bibnamefont
  {Pretko}},\ }\bibfield  {title} {\bibinfo {title} {Emergent gravity of
  fractons: {{Mach}}'s principle revisited},\ }\href
  {https://doi.org/10.1103/PhysRevD.96.024051} {\bibfield  {journal} {\bibinfo
  {journal} {Phys. Rev. D}\ }\textbf {\bibinfo {volume} {96}},\ \bibinfo
  {pages} {024051} (\bibinfo {year} {2017}{\natexlab{b}})}\BibitemShut
  {NoStop}%
\bibitem [{\citenamefont {Williamson}\ \emph {et~al.}(2019)\citenamefont
  {Williamson}, \citenamefont {Bi},\ and\ \citenamefont
  {Cheng}}]{williamson:2019}%
  \BibitemOpen
  \bibfield  {author} {\bibinfo {author} {\bibfnamefont {D.~J.}\ \bibnamefont
  {Williamson}}, \bibinfo {author} {\bibfnamefont {Z.}~\bibnamefont {Bi}},\
  and\ \bibinfo {author} {\bibfnamefont {M.}~\bibnamefont {Cheng}},\ }\bibfield
   {title} {\bibinfo {title} {Fractonic matter in symmetry-enriched {$U(1)$}
  gauge theory},\ }\href {https://doi.org/10.1103/PhysRevB.100.125150}
  {\bibfield  {journal} {\bibinfo  {journal} {Phys. Rev. B}\ }\textbf {\bibinfo
  {volume} {100}},\ \bibinfo {pages} {125150} (\bibinfo {year}
  {2019})}\BibitemShut {NoStop}%
\bibitem [{\citenamefont {Pretko}\ and\ \citenamefont
  {Radzihovsky}(2018{\natexlab{b}})}]{pretko:2018}%
  \BibitemOpen
  \bibfield  {author} {\bibinfo {author} {\bibfnamefont {M.}~\bibnamefont
  {Pretko}}\ and\ \bibinfo {author} {\bibfnamefont {L.}~\bibnamefont
  {Radzihovsky}},\ }\bibfield  {title} {\bibinfo {title} {Fracton-{{Elasticity
  Duality}}},\ }\href {https://doi.org/10.1103/PhysRevLett.120.195301}
  {\bibfield  {journal} {\bibinfo  {journal} {Phys. Rev. Lett.}\ }\textbf
  {\bibinfo {volume} {120}},\ \bibinfo {pages} {195301} (\bibinfo {year}
  {2018}{\natexlab{b}})}\BibitemShut {NoStop}%
\bibitem [{\citenamefont {Gromov}(2019)}]{gromov:2019}%
  \BibitemOpen
  \bibfield  {author} {\bibinfo {author} {\bibfnamefont {A.}~\bibnamefont
  {Gromov}},\ }\bibfield  {title} {\bibinfo {title} {Chiral {{Topological
  Elasticity}} and {{Fracton Order}}},\ }\href
  {https://doi.org/10.1103/PhysRevLett.122.076403} {\bibfield  {journal}
  {\bibinfo  {journal} {Phys. Rev. Lett.}\ }\textbf {\bibinfo {volume} {122}},\
  \bibinfo {pages} {076403} (\bibinfo {year} {2019})}\BibitemShut {NoStop}%
\bibitem [{\citenamefont {Kumar}\ and\ \citenamefont
  {Potter}(2019)}]{kumar:2019}%
  \BibitemOpen
  \bibfield  {author} {\bibinfo {author} {\bibfnamefont {A.}~\bibnamefont
  {Kumar}}\ and\ \bibinfo {author} {\bibfnamefont {A.~C.}\ \bibnamefont
  {Potter}},\ }\bibfield  {title} {\bibinfo {title} {Symmetry-enforced
  fractonicity and two-dimensional quantum crystal melting},\ }\href
  {https://doi.org/10.1103/PhysRevB.100.045119} {\bibfield  {journal} {\bibinfo
   {journal} {Phys. Rev. B}\ }\textbf {\bibinfo {volume} {100}},\ \bibinfo
  {pages} {045119} (\bibinfo {year} {2019})}\BibitemShut {NoStop}%
\bibitem [{\citenamefont {Zhai}\ and\ \citenamefont
  {Radzihovsky}(2019)}]{zhai:2019}%
  \BibitemOpen
  \bibfield  {author} {\bibinfo {author} {\bibfnamefont {Z.}~\bibnamefont
  {Zhai}}\ and\ \bibinfo {author} {\bibfnamefont {L.}~\bibnamefont
  {Radzihovsky}},\ }\bibfield  {title} {\bibinfo {title} {Two-dimensional
  melting via sine-{{Gordon}} duality},\ }\href
  {https://doi.org/10.1103/PhysRevB.100.094105} {\bibfield  {journal} {\bibinfo
   {journal} {Phys. Rev. B}\ }\textbf {\bibinfo {volume} {100}},\ \bibinfo
  {pages} {094105} (\bibinfo {year} {2019})}\BibitemShut {NoStop}%
\bibitem [{\citenamefont {Radzihovsky}(2020)}]{radzihovsky:2020a}%
  \BibitemOpen
  \bibfield  {author} {\bibinfo {author} {\bibfnamefont {L.}~\bibnamefont
  {Radzihovsky}},\ }\bibfield  {title} {\bibinfo {title} {Quantum {{Smectic
  Gauge Theory}}},\ }\href {https://doi.org/10.1103/PhysRevLett.125.267601}
  {\bibfield  {journal} {\bibinfo  {journal} {Phys. Rev. Lett.}\ }\textbf
  {\bibinfo {volume} {125}},\ \bibinfo {pages} {267601} (\bibinfo {year}
  {2020})}\BibitemShut {NoStop}%
\bibitem [{\citenamefont {Zhai}\ and\ \citenamefont
  {Radzihovsky}(2021)}]{zhai:2021}%
  \BibitemOpen
  \bibfield  {author} {\bibinfo {author} {\bibfnamefont {Z.}~\bibnamefont
  {Zhai}}\ and\ \bibinfo {author} {\bibfnamefont {L.}~\bibnamefont
  {Radzihovsky}},\ }\bibfield  {title} {\bibinfo {title} {Fractonic gauge
  theory of smectics},\ }\href {https://doi.org/10.1016/j.aop.2021.168509}
  {\bibfield  {journal} {\bibinfo  {journal} {Annals of Physics}\ }\bibinfo
  {series} {Special Issue on {{Philip W}}. {{Anderson}}},\ \textbf {\bibinfo
  {volume} {435}},\ \bibinfo {pages} {168509} (\bibinfo {year}
  {2021})}\BibitemShut {NoStop}%
\bibitem [{\citenamefont {Pretko}(2018)}]{pretko:2018c}%
  \BibitemOpen
  \bibfield  {author} {\bibinfo {author} {\bibfnamefont {M.}~\bibnamefont
  {Pretko}},\ }\bibfield  {title} {\bibinfo {title} {The fracton gauge
  principle},\ }\href {https://doi.org/10.1103/PhysRevB.98.115134} {\bibfield
  {journal} {\bibinfo  {journal} {Phys. Rev. B}\ }\textbf {\bibinfo {volume}
  {98}},\ \bibinfo {pages} {115134} (\bibinfo {year} {2018})}\BibitemShut
  {NoStop}%
\bibitem [{\citenamefont {Pretko}\ \emph {et~al.}(2019)\citenamefont {Pretko},
  \citenamefont {Zhai},\ and\ \citenamefont {Radzihovsky}}]{pretko:2019}%
  \BibitemOpen
  \bibfield  {author} {\bibinfo {author} {\bibfnamefont {M.}~\bibnamefont
  {Pretko}}, \bibinfo {author} {\bibfnamefont {Z.}~\bibnamefont {Zhai}},\ and\
  \bibinfo {author} {\bibfnamefont {L.}~\bibnamefont {Radzihovsky}},\
  }\bibfield  {title} {\bibinfo {title} {Crystal-to-fracton tensor gauge theory
  dualities},\ }\href {https://doi.org/10.1103/PhysRevB.100.134113} {\bibfield
  {journal} {\bibinfo  {journal} {Phys. Rev. B}\ }\textbf {\bibinfo {volume}
  {100}},\ \bibinfo {pages} {134113} (\bibinfo {year} {2019})}\BibitemShut
  {NoStop}%
\bibitem [{\citenamefont {Yuan}\ \emph {et~al.}(2020)\citenamefont {Yuan},
  \citenamefont {Chen},\ and\ \citenamefont {Ye}}]{yuan:2020}%
  \BibitemOpen
  \bibfield  {author} {\bibinfo {author} {\bibfnamefont {J.-K.}\ \bibnamefont
  {Yuan}}, \bibinfo {author} {\bibfnamefont {S.~A.}\ \bibnamefont {Chen}},\
  and\ \bibinfo {author} {\bibfnamefont {P.}~\bibnamefont {Ye}},\ }\bibfield
  {title} {\bibinfo {title} {Fractonic superfluids},\ }\href
  {https://doi.org/10.1103/PhysRevResearch.2.023267} {\bibfield  {journal}
  {\bibinfo  {journal} {Phys. Rev. Research}\ }\textbf {\bibinfo {volume}
  {2}},\ \bibinfo {pages} {023267} (\bibinfo {year} {2020})}\BibitemShut
  {NoStop}%
\bibitem [{\citenamefont {Chen}\ \emph {et~al.}(2021)\citenamefont {Chen},
  \citenamefont {Yuan},\ and\ \citenamefont {Ye}}]{chen:2021}%
  \BibitemOpen
  \bibfield  {author} {\bibinfo {author} {\bibfnamefont {S.~A.}\ \bibnamefont
  {Chen}}, \bibinfo {author} {\bibfnamefont {J.-K.}\ \bibnamefont {Yuan}},\
  and\ \bibinfo {author} {\bibfnamefont {P.}~\bibnamefont {Ye}},\ }\bibfield
  {title} {\bibinfo {title} {Fractonic superfluids. {{II}}. {{Condensing}}
  subdimensional particles},\ }\href
  {https://doi.org/10.1103/PhysRevResearch.3.013226} {\bibfield  {journal}
  {\bibinfo  {journal} {Phys. Rev. Res.}\ }\textbf {\bibinfo {volume} {3}},\
  \bibinfo {pages} {013226} (\bibinfo {year} {2021})}\BibitemShut {NoStop}%
\bibitem [{\citenamefont {Radzihovsky}(2022)}]{radzihovsky:2022}%
  \BibitemOpen
  \bibfield  {author} {\bibinfo {author} {\bibfnamefont {L.}~\bibnamefont
  {Radzihovsky}},\ }\bibfield  {title} {\bibinfo {title} {Lifshitz gauge
  duality},\ }\href {https://doi.org/10.1103/PhysRevB.106.224510} {\bibfield
  {journal} {\bibinfo  {journal} {Phys. Rev. B}\ }\textbf {\bibinfo {volume}
  {106}},\ \bibinfo {pages} {224510} (\bibinfo {year} {2022})}\BibitemShut
  {NoStop}%
\bibitem [{\citenamefont {Stahl}\ \emph {et~al.}(2023)\citenamefont {Stahl},
  \citenamefont {Qi}, \citenamefont {Glorioso}, \citenamefont {Lucas},\ and\
  \citenamefont {Nandkishore}}]{stahl:2023}%
  \BibitemOpen
  \bibfield  {author} {\bibinfo {author} {\bibfnamefont {C.}~\bibnamefont
  {Stahl}}, \bibinfo {author} {\bibfnamefont {M.}~\bibnamefont {Qi}}, \bibinfo
  {author} {\bibfnamefont {P.}~\bibnamefont {Glorioso}}, \bibinfo {author}
  {\bibfnamefont {A.}~\bibnamefont {Lucas}},\ and\ \bibinfo {author}
  {\bibfnamefont {R.}~\bibnamefont {Nandkishore}},\ }\href
  {https://doi.org/10.48550/arXiv.2303.09573} {\bibinfo {title} {Fracton
  superfluid hydrodynamics}} (\bibinfo {year} {2023}),\ \Eprint
  {https://arxiv.org/abs/2303.09573} {arxiv:2303.09573} \BibitemShut {NoStop}%
\bibitem [{\citenamefont {Lake}\ \emph {et~al.}(2022)\citenamefont {Lake},
  \citenamefont {Hermele},\ and\ \citenamefont {Senthil}}]{lake:2022}%
  \BibitemOpen
  \bibfield  {author} {\bibinfo {author} {\bibfnamefont {E.}~\bibnamefont
  {Lake}}, \bibinfo {author} {\bibfnamefont {M.}~\bibnamefont {Hermele}},\ and\
  \bibinfo {author} {\bibfnamefont {T.}~\bibnamefont {Senthil}},\ }\bibfield
  {title} {\bibinfo {title} {Dipolar {{Bose-Hubbard}} model},\ }\href
  {https://doi.org/10.1103/PhysRevB.106.064511} {\bibfield  {journal} {\bibinfo
   {journal} {Phys. Rev. B}\ }\textbf {\bibinfo {volume} {106}},\ \bibinfo
  {pages} {064511} (\bibinfo {year} {2022})}\BibitemShut {NoStop}%
\bibitem [{\citenamefont {Zechmann}\ \emph {et~al.}(2023)\citenamefont
  {Zechmann}, \citenamefont {Altman}, \citenamefont {Knap},\ and\ \citenamefont
  {Feldmeier}}]{zechmann:2023}%
  \BibitemOpen
  \bibfield  {author} {\bibinfo {author} {\bibfnamefont {P.}~\bibnamefont
  {Zechmann}}, \bibinfo {author} {\bibfnamefont {E.}~\bibnamefont {Altman}},
  \bibinfo {author} {\bibfnamefont {M.}~\bibnamefont {Knap}},\ and\ \bibinfo
  {author} {\bibfnamefont {J.}~\bibnamefont {Feldmeier}},\ }\bibfield  {title}
  {\bibinfo {title} {Fractonic {{Luttinger}} liquids and supersolids in a
  constrained {{Bose-Hubbard}} model},\ }\href
  {https://doi.org/10.1103/PhysRevB.107.195131} {\bibfield  {journal} {\bibinfo
   {journal} {Phys. Rev. B}\ }\textbf {\bibinfo {volume} {107}},\ \bibinfo
  {pages} {195131} (\bibinfo {year} {2023})}\BibitemShut {NoStop}%
\bibitem [{\citenamefont {Lake}\ \emph {et~al.}(2023)\citenamefont {Lake},
  \citenamefont {Lee}, \citenamefont {Han},\ and\ \citenamefont
  {Senthil}}]{lake:2023}%
  \BibitemOpen
  \bibfield  {author} {\bibinfo {author} {\bibfnamefont {E.}~\bibnamefont
  {Lake}}, \bibinfo {author} {\bibfnamefont {H.-Y.}\ \bibnamefont {Lee}},
  \bibinfo {author} {\bibfnamefont {J.~H.}\ \bibnamefont {Han}},\ and\ \bibinfo
  {author} {\bibfnamefont {T.}~\bibnamefont {Senthil}},\ }\bibfield  {title}
  {\bibinfo {title} {Dipole condensates in tilted {{Bose-Hubbard}} chains},\
  }\href {https://doi.org/10.1103/PhysRevB.107.195132} {\bibfield  {journal}
  {\bibinfo  {journal} {Phys. Rev. B}\ }\textbf {\bibinfo {volume} {107}},\
  \bibinfo {pages} {195132} (\bibinfo {year} {2023})}\BibitemShut {NoStop}%
\bibitem [{\citenamefont {Lake}\ and\ \citenamefont
  {Senthil}(2023)}]{lake:2023a}%
  \BibitemOpen
  \bibfield  {author} {\bibinfo {author} {\bibfnamefont {E.}~\bibnamefont
  {Lake}}\ and\ \bibinfo {author} {\bibfnamefont {T.}~\bibnamefont {Senthil}},\
  }\bibfield  {title} {\bibinfo {title} {Non-{{Fermi Liquids}} from {{Kinetic
  Constraints}} in {{Tilted Optical Lattices}}},\ }\href
  {https://doi.org/10.1103/PhysRevLett.131.043403} {\bibfield  {journal}
  {\bibinfo  {journal} {Phys. Rev. Lett.}\ }\textbf {\bibinfo {volume} {131}},\
  \bibinfo {pages} {043403} (\bibinfo {year} {2023})}\BibitemShut {NoStop}%
\bibitem [{\citenamefont {{Guardado-Sanchez}}\ \emph
  {et~al.}(2020)\citenamefont {{Guardado-Sanchez}}, \citenamefont
  {Morningstar}, \citenamefont {Spar}, \citenamefont {Brown}, \citenamefont
  {Huse},\ and\ \citenamefont {Bakr}}]{guardado-sanchez:2020}%
  \BibitemOpen
  \bibfield  {author} {\bibinfo {author} {\bibfnamefont {E.}~\bibnamefont
  {{Guardado-Sanchez}}}, \bibinfo {author} {\bibfnamefont {A.}~\bibnamefont
  {Morningstar}}, \bibinfo {author} {\bibfnamefont {B.~M.}\ \bibnamefont
  {Spar}}, \bibinfo {author} {\bibfnamefont {P.~T.}\ \bibnamefont {Brown}},
  \bibinfo {author} {\bibfnamefont {D.~A.}\ \bibnamefont {Huse}},\ and\
  \bibinfo {author} {\bibfnamefont {W.~S.}\ \bibnamefont {Bakr}},\ }\bibfield
  {title} {\bibinfo {title} {Subdiffusion and heat transport in a tilted
  two-dimensional {{Fermi-Hubbard}} system},\ }\href
  {https://doi.org/10.1103/PhysRevX.10.011042} {\bibfield  {journal} {\bibinfo
  {journal} {Phys. Rev. X}\ }\textbf {\bibinfo {volume} {10}},\ \bibinfo
  {pages} {011042} (\bibinfo {year} {2020})}\BibitemShut {NoStop}%
\bibitem [{\citenamefont {Scherg}\ \emph {et~al.}(2021)\citenamefont {Scherg},
  \citenamefont {Kohlert}, \citenamefont {Sala}, \citenamefont {Pollmann},
  \citenamefont {Hebbe~Madhusudhana}, \citenamefont {Bloch},\ and\
  \citenamefont {Aidelsburger}}]{scherg:2021}%
  \BibitemOpen
  \bibfield  {author} {\bibinfo {author} {\bibfnamefont {S.}~\bibnamefont
  {Scherg}}, \bibinfo {author} {\bibfnamefont {T.}~\bibnamefont {Kohlert}},
  \bibinfo {author} {\bibfnamefont {P.}~\bibnamefont {Sala}}, \bibinfo {author}
  {\bibfnamefont {F.}~\bibnamefont {Pollmann}}, \bibinfo {author}
  {\bibfnamefont {B.}~\bibnamefont {Hebbe~Madhusudhana}}, \bibinfo {author}
  {\bibfnamefont {I.}~\bibnamefont {Bloch}},\ and\ \bibinfo {author}
  {\bibfnamefont {M.}~\bibnamefont {Aidelsburger}},\ }\bibfield  {title}
  {\bibinfo {title} {Observing non-ergodicity due to kinetic constraints in
  tilted {{Fermi-Hubbard}} chains},\ }\href
  {https://doi.org/10.1038/s41467-021-24726-0} {\bibfield  {journal} {\bibinfo
  {journal} {Nat Commun}\ }\textbf {\bibinfo {volume} {12}},\ \bibinfo {pages}
  {4490} (\bibinfo {year} {2021})}\BibitemShut {NoStop}%
\bibitem [{\citenamefont {Zahn}\ \emph {et~al.}(2022)\citenamefont {Zahn},
  \citenamefont {Singh}, \citenamefont {Kosch}, \citenamefont {Asteria},
  \citenamefont {Freystatzky}, \citenamefont {Sengstock}, \citenamefont
  {Mathey},\ and\ \citenamefont {Weitenberg}}]{zahn:2022}%
  \BibitemOpen
  \bibfield  {author} {\bibinfo {author} {\bibfnamefont {H.~P.}\ \bibnamefont
  {Zahn}}, \bibinfo {author} {\bibfnamefont {V.~P.}\ \bibnamefont {Singh}},
  \bibinfo {author} {\bibfnamefont {M.~N.}\ \bibnamefont {Kosch}}, \bibinfo
  {author} {\bibfnamefont {L.}~\bibnamefont {Asteria}}, \bibinfo {author}
  {\bibfnamefont {L.}~\bibnamefont {Freystatzky}}, \bibinfo {author}
  {\bibfnamefont {K.}~\bibnamefont {Sengstock}}, \bibinfo {author}
  {\bibfnamefont {L.}~\bibnamefont {Mathey}},\ and\ \bibinfo {author}
  {\bibfnamefont {C.}~\bibnamefont {Weitenberg}},\ }\bibfield  {title}
  {\bibinfo {title} {Formation of spontaneous density-wave patterns in dc
  driven lattices},\ }\href {https://doi.org/10.1103/PhysRevX.12.021014}
  {\bibfield  {journal} {\bibinfo  {journal} {Phys. Rev. X}\ }\textbf {\bibinfo
  {volume} {12}},\ \bibinfo {pages} {021014} (\bibinfo {year}
  {2022})}\BibitemShut {NoStop}%
\bibitem [{\citenamefont {Kohlert}\ \emph {et~al.}(2023)\citenamefont
  {Kohlert}, \citenamefont {Scherg}, \citenamefont {Sala}, \citenamefont
  {Pollmann}, \citenamefont {Hebbe~Madhusudhana}, \citenamefont {Bloch},\ and\
  \citenamefont {Aidelsburger}}]{kohlert:2023}%
  \BibitemOpen
  \bibfield  {author} {\bibinfo {author} {\bibfnamefont {T.}~\bibnamefont
  {Kohlert}}, \bibinfo {author} {\bibfnamefont {S.}~\bibnamefont {Scherg}},
  \bibinfo {author} {\bibfnamefont {P.}~\bibnamefont {Sala}}, \bibinfo {author}
  {\bibfnamefont {F.}~\bibnamefont {Pollmann}}, \bibinfo {author}
  {\bibfnamefont {B.}~\bibnamefont {Hebbe~Madhusudhana}}, \bibinfo {author}
  {\bibfnamefont {I.}~\bibnamefont {Bloch}},\ and\ \bibinfo {author}
  {\bibfnamefont {M.}~\bibnamefont {Aidelsburger}},\ }\bibfield  {title}
  {\bibinfo {title} {Exploring the {{Regime}} of {{Fragmentation}} in
  {{Strongly Tilted Fermi-Hubbard Chains}}},\ }\href
  {https://doi.org/10.1103/PhysRevLett.130.010201} {\bibfield  {journal}
  {\bibinfo  {journal} {Phys. Rev. Lett.}\ }\textbf {\bibinfo {volume} {130}},\
  \bibinfo {pages} {010201} (\bibinfo {year} {2023})}\BibitemShut {NoStop}%
\bibitem [{\citenamefont {Moudgalya}\ \emph {et~al.}(2021)\citenamefont
  {Moudgalya}, \citenamefont {Prem}, \citenamefont {Huse},\ and\ \citenamefont
  {Chan}}]{moudgalya:2021}%
  \BibitemOpen
  \bibfield  {author} {\bibinfo {author} {\bibfnamefont {S.}~\bibnamefont
  {Moudgalya}}, \bibinfo {author} {\bibfnamefont {A.}~\bibnamefont {Prem}},
  \bibinfo {author} {\bibfnamefont {D.~A.}\ \bibnamefont {Huse}},\ and\
  \bibinfo {author} {\bibfnamefont {A.}~\bibnamefont {Chan}},\ }\bibfield
  {title} {\bibinfo {title} {Spectral statistics in constrained many-body
  quantum chaotic systems},\ }\href
  {https://doi.org/10.1103/PhysRevResearch.3.023176} {\bibfield  {journal}
  {\bibinfo  {journal} {Phys. Rev. Res.}\ }\textbf {\bibinfo {volume} {3}},\
  \bibinfo {pages} {023176} (\bibinfo {year} {2021})}\BibitemShut {NoStop}%
\bibitem [{\citenamefont {Feng}\ and\ \citenamefont
  {Skinner}(2022)}]{feng:2022}%
  \BibitemOpen
  \bibfield  {author} {\bibinfo {author} {\bibfnamefont {X.}~\bibnamefont
  {Feng}}\ and\ \bibinfo {author} {\bibfnamefont {B.}~\bibnamefont {Skinner}},\
  }\bibfield  {title} {\bibinfo {title} {Hilbert space fragmentation produces
  an effective attraction between fractons},\ }\href
  {https://doi.org/10.1103/PhysRevResearch.4.013053} {\bibfield  {journal}
  {\bibinfo  {journal} {Phys. Rev. Res.}\ }\textbf {\bibinfo {volume} {4}},\
  \bibinfo {pages} {013053} (\bibinfo {year} {2022})}\BibitemShut {NoStop}%
\bibitem [{\citenamefont {Giamarchi}(2004)}]{giamarchi:2004}%
  \BibitemOpen
  \bibfield  {author} {\bibinfo {author} {\bibfnamefont {T.}~\bibnamefont
  {Giamarchi}},\ }\href
  {https://doi.org/10.1093/acprof:oso/9780198525004.001.0001} {\emph {\bibinfo
  {title} {Quantum Physics in One Dimension}}},\ \bibinfo {series} {The
  International Series of Monographs on Physics}\ No.\ \bibinfo {number} {121}\
  (\bibinfo  {publisher} {{Oxford University Press}},\ \bibinfo {address}
  {{Oxford}},\ \bibinfo {year} {2004})\BibitemShut {NoStop}%
\bibitem [{\citenamefont {Schwabl}(2004)}]{schwabl:2004}%
  \BibitemOpen
  \bibfield  {author} {\bibinfo {author} {\bibfnamefont {F.}~\bibnamefont
  {Schwabl}},\ }\href@noop {} {\emph {\bibinfo {title} {Advanced {{Quantum
  Mechanics}}}}},\ \bibinfo {edition} {second edition}\ ed.\ (\bibinfo
  {publisher} {{Springer Berlin Heidelberg}},\ \bibinfo {address} {{Berlin,
  Heidelberg}},\ \bibinfo {year} {2004})\BibitemShut {NoStop}%
\bibitem [{\citenamefont {Freericks}\ \emph {et~al.}(2013)\citenamefont
  {Freericks}, \citenamefont {Turkowski}, \citenamefont {Krishnamurthy},\ and\
  \citenamefont {Knap}}]{freericks:2013}%
  \BibitemOpen
  \bibfield  {author} {\bibinfo {author} {\bibfnamefont {J.~K.}\ \bibnamefont
  {Freericks}}, \bibinfo {author} {\bibfnamefont {V.}~\bibnamefont
  {Turkowski}}, \bibinfo {author} {\bibfnamefont {H.~R.}\ \bibnamefont
  {Krishnamurthy}},\ and\ \bibinfo {author} {\bibfnamefont {M.}~\bibnamefont
  {Knap}},\ }\bibfield  {title} {\bibinfo {title} {Spectral moment sum rules
  for the retarded {{Green}}'s function and self-energy of the inhomogeneous
  {{Bose-Hubbard}} model in equilibrium and nonequilibrium},\ }\href
  {https://doi.org/10.1103/PhysRevA.87.013628} {\bibfield  {journal} {\bibinfo
  {journal} {Phys. Rev. A}\ }\textbf {\bibinfo {volume} {87}},\ \bibinfo
  {pages} {013628} (\bibinfo {year} {2013})}\BibitemShut {NoStop}%
\bibitem [{\citenamefont {White}(1992)}]{white:1992}%
  \BibitemOpen
  \bibfield  {author} {\bibinfo {author} {\bibfnamefont {S.~R.}\ \bibnamefont
  {White}},\ }\bibfield  {title} {\bibinfo {title} {Density matrix formulation
  for quantum renormalization groups},\ }\href
  {https://doi.org/10.1103/PhysRevLett.69.2863} {\bibfield  {journal} {\bibinfo
   {journal} {Phys. Rev. Lett.}\ }\textbf {\bibinfo {volume} {69}},\ \bibinfo
  {pages} {2863} (\bibinfo {year} {1992})}\BibitemShut {NoStop}%
\bibitem [{\citenamefont {Vidal}(2007)}]{vidal:2007}%
  \BibitemOpen
  \bibfield  {author} {\bibinfo {author} {\bibfnamefont {G.}~\bibnamefont
  {Vidal}},\ }\bibfield  {title} {\bibinfo {title} {Classical {{Simulation}} of
  {{Infinite-Size Quantum Lattice Systems}} in {{One Spatial Dimension}}},\
  }\href {https://doi.org/10.1103/PhysRevLett.98.070201} {\bibfield  {journal}
  {\bibinfo  {journal} {Phys. Rev. Lett.}\ }\textbf {\bibinfo {volume} {98}},\
  \bibinfo {pages} {070201} (\bibinfo {year} {2007})}\BibitemShut {NoStop}%
\bibitem [{\citenamefont {Singh}\ \emph {et~al.}(2010)\citenamefont {Singh},
  \citenamefont {Pfeifer},\ and\ \citenamefont {Vidal}}]{singh:2010}%
  \BibitemOpen
  \bibfield  {author} {\bibinfo {author} {\bibfnamefont {S.}~\bibnamefont
  {Singh}}, \bibinfo {author} {\bibfnamefont {R.~N.~C.}\ \bibnamefont
  {Pfeifer}},\ and\ \bibinfo {author} {\bibfnamefont {G.}~\bibnamefont
  {Vidal}},\ }\bibfield  {title} {\bibinfo {title} {Tensor network
  decompositions in the presence of a global symmetry},\ }\href
  {https://doi.org/10.1103/PhysRevA.82.050301} {\bibfield  {journal} {\bibinfo
  {journal} {Phys. Rev. A}\ }\textbf {\bibinfo {volume} {82}},\ \bibinfo
  {pages} {050301} (\bibinfo {year} {2010})}\BibitemShut {NoStop}%
\bibitem [{\citenamefont {Singh}\ \emph {et~al.}(2011)\citenamefont {Singh},
  \citenamefont {Pfeifer},\ and\ \citenamefont {Vidal}}]{singh:2011}%
  \BibitemOpen
  \bibfield  {author} {\bibinfo {author} {\bibfnamefont {S.}~\bibnamefont
  {Singh}}, \bibinfo {author} {\bibfnamefont {R.~N.~C.}\ \bibnamefont
  {Pfeifer}},\ and\ \bibinfo {author} {\bibfnamefont {G.}~\bibnamefont
  {Vidal}},\ }\bibfield  {title} {\bibinfo {title} {Tensor network states and
  algorithms in the presence of a global {{U}}(1) symmetry},\ }\href
  {https://doi.org/10.1103/PhysRevB.83.115125} {\bibfield  {journal} {\bibinfo
  {journal} {Phys. Rev. B}\ }\textbf {\bibinfo {volume} {83}},\ \bibinfo
  {pages} {115125} (\bibinfo {year} {2011})}\BibitemShut {NoStop}%
\bibitem [{\citenamefont {Zaletel}\ \emph {et~al.}(2013)\citenamefont
  {Zaletel}, \citenamefont {Mong},\ and\ \citenamefont
  {Pollmann}}]{zaletel:2013}%
  \BibitemOpen
  \bibfield  {author} {\bibinfo {author} {\bibfnamefont {M.~P.}\ \bibnamefont
  {Zaletel}}, \bibinfo {author} {\bibfnamefont {R.~S.~K.}\ \bibnamefont
  {Mong}},\ and\ \bibinfo {author} {\bibfnamefont {F.}~\bibnamefont
  {Pollmann}},\ }\bibfield  {title} {\bibinfo {title} {Topological
  characterization of fractional quantum hall ground states from microscopic
  hamiltonians},\ }\href {https://doi.org/10.1103/PhysRevLett.110.236801}
  {\bibfield  {journal} {\bibinfo  {journal} {Phys. Rev. Lett.}\ }\textbf
  {\bibinfo {volume} {110}},\ \bibinfo {pages} {236801} (\bibinfo {year}
  {2013})}\BibitemShut {NoStop}%
\bibitem [{\citenamefont {White}(2005)}]{white:2005}%
  \BibitemOpen
  \bibfield  {author} {\bibinfo {author} {\bibfnamefont {S.~R.}\ \bibnamefont
  {White}},\ }\bibfield  {title} {\bibinfo {title} {Density matrix
  renormalization group algorithms with a single center site},\ }\href
  {https://doi.org/10.1103/PhysRevB.72.180403} {\bibfield  {journal} {\bibinfo
  {journal} {Phys. Rev. B}\ }\textbf {\bibinfo {volume} {72}},\ \bibinfo
  {pages} {180403} (\bibinfo {year} {2005})}\BibitemShut {NoStop}%
\bibitem [{\citenamefont {Hubig}\ \emph {et~al.}(2015)\citenamefont {Hubig},
  \citenamefont {McCulloch}, \citenamefont {Schollw{\"o}ck},\ and\
  \citenamefont {Wolf}}]{hubig:2015}%
  \BibitemOpen
  \bibfield  {author} {\bibinfo {author} {\bibfnamefont {C.}~\bibnamefont
  {Hubig}}, \bibinfo {author} {\bibfnamefont {I.~P.}\ \bibnamefont
  {McCulloch}}, \bibinfo {author} {\bibfnamefont {U.}~\bibnamefont
  {Schollw{\"o}ck}},\ and\ \bibinfo {author} {\bibfnamefont {F.~A.}\
  \bibnamefont {Wolf}},\ }\bibfield  {title} {\bibinfo {title} {Strictly
  single-site {{DMRG}} algorithm with subspace expansion},\ }\href
  {https://doi.org/10.1103/PhysRevB.91.155115} {\bibfield  {journal} {\bibinfo
  {journal} {Phys. Rev. B}\ }\textbf {\bibinfo {volume} {91}},\ \bibinfo
  {pages} {155115} (\bibinfo {year} {2015})}\BibitemShut {NoStop}%
\bibitem [{\citenamefont {Haegeman}\ \emph {et~al.}(2011)\citenamefont
  {Haegeman}, \citenamefont {Cirac}, \citenamefont {Osborne}, \citenamefont
  {Pi{\v z}orn}, \citenamefont {Verschelde},\ and\ \citenamefont
  {Verstraete}}]{haegeman:2011}%
  \BibitemOpen
  \bibfield  {author} {\bibinfo {author} {\bibfnamefont {J.}~\bibnamefont
  {Haegeman}}, \bibinfo {author} {\bibfnamefont {J.~I.}\ \bibnamefont {Cirac}},
  \bibinfo {author} {\bibfnamefont {T.~J.}\ \bibnamefont {Osborne}}, \bibinfo
  {author} {\bibfnamefont {I.}~\bibnamefont {Pi{\v z}orn}}, \bibinfo {author}
  {\bibfnamefont {H.}~\bibnamefont {Verschelde}},\ and\ \bibinfo {author}
  {\bibfnamefont {F.}~\bibnamefont {Verstraete}},\ }\bibfield  {title}
  {\bibinfo {title} {Time-{{Dependent Variational Principle}} for {{Quantum
  Lattices}}},\ }\href {https://doi.org/10.1103/PhysRevLett.107.070601}
  {\bibfield  {journal} {\bibinfo  {journal} {Phys. Rev. Lett.}\ }\textbf
  {\bibinfo {volume} {107}},\ \bibinfo {pages} {070601} (\bibinfo {year}
  {2011})}\BibitemShut {NoStop}%
\bibitem [{\citenamefont {Haegeman}\ \emph {et~al.}(2016)\citenamefont
  {Haegeman}, \citenamefont {Lubich}, \citenamefont {Oseledets}, \citenamefont
  {Vandereycken},\ and\ \citenamefont {Verstraete}}]{haegeman:2016}%
  \BibitemOpen
  \bibfield  {author} {\bibinfo {author} {\bibfnamefont {J.}~\bibnamefont
  {Haegeman}}, \bibinfo {author} {\bibfnamefont {C.}~\bibnamefont {Lubich}},
  \bibinfo {author} {\bibfnamefont {I.}~\bibnamefont {Oseledets}}, \bibinfo
  {author} {\bibfnamefont {B.}~\bibnamefont {Vandereycken}},\ and\ \bibinfo
  {author} {\bibfnamefont {F.}~\bibnamefont {Verstraete}},\ }\bibfield  {title}
  {\bibinfo {title} {Unifying time evolution and optimization with matrix
  product states},\ }\href {https://doi.org/10.1103/PhysRevB.94.165116}
  {\bibfield  {journal} {\bibinfo  {journal} {Phys. Rev. B}\ }\textbf {\bibinfo
  {volume} {94}},\ \bibinfo {pages} {165116} (\bibinfo {year}
  {2016})}\BibitemShut {NoStop}%
\bibitem [{\citenamefont {Phien}\ \emph {et~al.}(2012)\citenamefont {Phien},
  \citenamefont {Vidal},\ and\ \citenamefont {McCulloch}}]{phien:2012}%
  \BibitemOpen
  \bibfield  {author} {\bibinfo {author} {\bibfnamefont {H.~N.}\ \bibnamefont
  {Phien}}, \bibinfo {author} {\bibfnamefont {G.}~\bibnamefont {Vidal}},\ and\
  \bibinfo {author} {\bibfnamefont {I.~P.}\ \bibnamefont {McCulloch}},\
  }\bibfield  {title} {\bibinfo {title} {Infinite boundary conditions for
  matrix product state calculations},\ }\href
  {https://doi.org/10.1103/PhysRevB.86.245107} {\bibfield  {journal} {\bibinfo
  {journal} {Phys. Rev. B}\ }\textbf {\bibinfo {volume} {86}},\ \bibinfo
  {pages} {245107} (\bibinfo {year} {2012})}\BibitemShut {NoStop}%
\bibitem [{\citenamefont {Milsted}\ \emph {et~al.}(2013)\citenamefont
  {Milsted}, \citenamefont {Haegeman}, \citenamefont {Osborne},\ and\
  \citenamefont {Verstraete}}]{milsted:2013}%
  \BibitemOpen
  \bibfield  {author} {\bibinfo {author} {\bibfnamefont {A.}~\bibnamefont
  {Milsted}}, \bibinfo {author} {\bibfnamefont {J.}~\bibnamefont {Haegeman}},
  \bibinfo {author} {\bibfnamefont {T.~J.}\ \bibnamefont {Osborne}},\ and\
  \bibinfo {author} {\bibfnamefont {F.}~\bibnamefont {Verstraete}},\ }\bibfield
   {title} {\bibinfo {title} {Variational matrix product ansatz for nonuniform
  dynamics in the thermodynamic limit},\ }\href
  {https://doi.org/10.1103/PhysRevB.88.155116} {\bibfield  {journal} {\bibinfo
  {journal} {Phys. Rev. B}\ }\textbf {\bibinfo {volume} {88}},\ \bibinfo
  {pages} {155116} (\bibinfo {year} {2013})}\BibitemShut {NoStop}%
\bibitem [{\citenamefont {Press}(1992)}]{press:1992}%
  \BibitemOpen
  \bibinfo {editor} {\bibfnamefont {W.~H.}\ \bibnamefont {Press}},\ ed.,\
  \href@noop {} {\emph {\bibinfo {title} {Numerical Recipes in {{C}}: The Art
  of Scientific Computing}}},\ \bibinfo {edition} {2nd}\ ed.\ (\bibinfo
  {publisher} {{Cambridge University Press}},\ \bibinfo {address} {{Cambridge ;
  New York}},\ \bibinfo {year} {1992})\BibitemShut {NoStop}%
\bibitem [{\citenamefont {White}\ and\ \citenamefont
  {Affleck}(2008)}]{white:2008}%
  \BibitemOpen
  \bibfield  {author} {\bibinfo {author} {\bibfnamefont {S.~R.}\ \bibnamefont
  {White}}\ and\ \bibinfo {author} {\bibfnamefont {I.}~\bibnamefont
  {Affleck}},\ }\bibfield  {title} {\bibinfo {title} {Spectral function for the
  {$S=1$} {{Heisenberg}} antiferromagetic chain},\ }\href
  {https://doi.org/10.1103/PhysRevB.77.134437} {\bibfield  {journal} {\bibinfo
  {journal} {Phys. Rev. B}\ }\textbf {\bibinfo {volume} {77}},\ \bibinfo
  {pages} {134437} (\bibinfo {year} {2008})}\BibitemShut {NoStop}%
\bibitem [{\citenamefont {Haegeman}\ \emph {et~al.}(2013)\citenamefont
  {Haegeman}, \citenamefont {Michalakis}, \citenamefont {Nachtergaele},
  \citenamefont {Osborne}, \citenamefont {Schuch},\ and\ \citenamefont
  {Verstraete}}]{haegeman:2013}%
  \BibitemOpen
  \bibfield  {author} {\bibinfo {author} {\bibfnamefont {J.}~\bibnamefont
  {Haegeman}}, \bibinfo {author} {\bibfnamefont {S.}~\bibnamefont
  {Michalakis}}, \bibinfo {author} {\bibfnamefont {B.}~\bibnamefont
  {Nachtergaele}}, \bibinfo {author} {\bibfnamefont {T.~J.}\ \bibnamefont
  {Osborne}}, \bibinfo {author} {\bibfnamefont {N.}~\bibnamefont {Schuch}},\
  and\ \bibinfo {author} {\bibfnamefont {F.}~\bibnamefont {Verstraete}},\
  }\bibfield  {title} {\bibinfo {title} {Elementary {{Excitations}} in {{Gapped
  Quantum Spin Systems}}},\ }\href
  {https://doi.org/10.1103/PhysRevLett.111.080401} {\bibfield  {journal}
  {\bibinfo  {journal} {Phys. Rev. Lett.}\ }\textbf {\bibinfo {volume} {111}},\
  \bibinfo {pages} {080401} (\bibinfo {year} {2013})}\BibitemShut {NoStop}%
\bibitem [{\citenamefont {Vanderstraeten}\ \emph {et~al.}(2019)\citenamefont
  {Vanderstraeten}, \citenamefont {Haegeman},\ and\ \citenamefont
  {Verstraete}}]{vanderstraeten:2019}%
  \BibitemOpen
  \bibfield  {author} {\bibinfo {author} {\bibfnamefont {L.}~\bibnamefont
  {Vanderstraeten}}, \bibinfo {author} {\bibfnamefont {J.}~\bibnamefont
  {Haegeman}},\ and\ \bibinfo {author} {\bibfnamefont {F.}~\bibnamefont
  {Verstraete}},\ }\bibfield  {title} {\bibinfo {title} {Tangent-space methods
  for uniform matrix product states},\ }\href
  {https://doi.org/10.21468/SciPostPhysLectNotes.7} {\bibfield  {journal}
  {\bibinfo  {journal} {SciPost Phys. Lect. Notes}\ ,\ \bibinfo {pages} {007}}
  (\bibinfo {year} {2019})}\BibitemShut {NoStop}%
\bibitem [{\citenamefont {{van Oosten}}\ \emph {et~al.}(2001)\citenamefont
  {{van Oosten}}, \citenamefont {{van der Straten}},\ and\ \citenamefont
  {Stoof}}]{vanoosten:2001}%
  \BibitemOpen
  \bibfield  {author} {\bibinfo {author} {\bibfnamefont {D.}~\bibnamefont {{van
  Oosten}}}, \bibinfo {author} {\bibfnamefont {P.}~\bibnamefont {{van der
  Straten}}},\ and\ \bibinfo {author} {\bibfnamefont {H.~T.~C.}\ \bibnamefont
  {Stoof}},\ }\bibfield  {title} {\bibinfo {title} {Quantum phases in an
  optical lattice},\ }\href {https://doi.org/10.1103/PhysRevA.63.053601}
  {\bibfield  {journal} {\bibinfo  {journal} {Phys. Rev. A}\ }\textbf {\bibinfo
  {volume} {63}},\ \bibinfo {pages} {053601} (\bibinfo {year}
  {2001})}\BibitemShut {NoStop}%
\bibitem [{\citenamefont {Seidel}\ \emph {et~al.}(2005)\citenamefont {Seidel},
  \citenamefont {Fu}, \citenamefont {Lee}, \citenamefont {Leinaas},\ and\
  \citenamefont {Moore}}]{seidel:2005}%
  \BibitemOpen
  \bibfield  {author} {\bibinfo {author} {\bibfnamefont {A.}~\bibnamefont
  {Seidel}}, \bibinfo {author} {\bibfnamefont {H.}~\bibnamefont {Fu}}, \bibinfo
  {author} {\bibfnamefont {D.-H.}\ \bibnamefont {Lee}}, \bibinfo {author}
  {\bibfnamefont {J.~M.}\ \bibnamefont {Leinaas}},\ and\ \bibinfo {author}
  {\bibfnamefont {J.}~\bibnamefont {Moore}},\ }\bibfield  {title} {\bibinfo
  {title} {Incompressible quantum liquids and new conservation laws},\ }\href
  {https://doi.org/10.1103/PhysRevLett.95.266405} {\bibfield  {journal}
  {\bibinfo  {journal} {Phys. Rev. Lett.}\ }\textbf {\bibinfo {volume} {95}},\
  \bibinfo {pages} {266405} (\bibinfo {year} {2005})}\BibitemShut {NoStop}%
\bibitem [{\citenamefont {Boesl}\ \emph {et~al.}(2023)\citenamefont {Boesl},
  \citenamefont {Zechmann}, \citenamefont {Feldmeier},\ and\ \citenamefont
  {Knap}}]{boesl:2023}%
  \BibitemOpen
  \bibfield  {author} {\bibinfo {author} {\bibfnamefont {J.}~\bibnamefont
  {Boesl}}, \bibinfo {author} {\bibfnamefont {P.}~\bibnamefont {Zechmann}},
  \bibinfo {author} {\bibfnamefont {J.}~\bibnamefont {Feldmeier}},\ and\
  \bibinfo {author} {\bibfnamefont {M.}~\bibnamefont {Knap}},\ }\href@noop {}
  {\bibinfo {title} {in preparation}} (\bibinfo {year} {2023})\BibitemShut
  {NoStop}%
\bibitem [{\citenamefont {Hauschild}\ and\ \citenamefont
  {Pollmann}(2018)}]{hauschild:2018}%
  \BibitemOpen
  \bibfield  {author} {\bibinfo {author} {\bibfnamefont {J.}~\bibnamefont
  {Hauschild}}\ and\ \bibinfo {author} {\bibfnamefont {F.}~\bibnamefont
  {Pollmann}},\ }\bibfield  {title} {\bibinfo {title} {Efficient numerical
  simulations with tensor networks: {{Tensor Network Python}} ({{TeNPy}})},\
  }\href {https://doi.org/10.21468/SciPostPhysLectNotes.5} {\bibfield
  {journal} {\bibinfo  {journal} {SciPost Phys. Lect. Notes}\ ,\ \bibinfo
  {pages} {5}} (\bibinfo {year} {2018})}\BibitemShut {NoStop}%
\bibitem [{zen()}]{zenodo}%
  \BibitemOpen
  \href@noop {} {\bibinfo {title} {All data and simulation codes are available
  upon reasonable request at
  \href{https://doi.org/10.5281/zenodo.10014288}{10.5281/zenodo.10014288}}}\BibitemShut
  {NoStop}%
\end{thebibliography}%

\end{document}